\documentclass[journal,twocolumn]{IEEEtran}
\IEEEoverridecommandlockouts

\usepackage{geometry}
\usepackage[noend]{algpseudocode}
\usepackage{balance}
\usepackage{algorithmicx,algorithm}
\usepackage{eqparbox}
\usepackage{amsmath}
\usepackage{subfigure}
\usepackage{verbatim}
\usepackage{braket}
\usepackage{cases}
\usepackage{amsthm,amsmath,amssymb,bm,bbm}
\allowdisplaybreaks[4]

\newtheorem{lemma}{Lemma}
\newtheorem{theorem}{Theorem}
\newtheorem{remark}{Remark}

\usepackage{mathrsfs}

\usepackage{hyperref}
\hypersetup{hidelinks, 
colorlinks=true,
allcolors=black,
pdfstartview=Fit,
breaklinks=true}

\usepackage{graphicx}
\usepackage{epstopdf}
\usepackage[numbers,sort&compress]{natbib}

\begin{document}
%
%
%
%


\title{Mutual Information for Electromagnetic Information Theory Based on Random Fields
}
	\author{
		\IEEEauthorblockN{
		Zhongzhichao~Wan,~\IEEEmembership{Student Member,~IEEE,}
		Jieao~Zhu,~\IEEEmembership{Student Member,~IEEE,}
		\\
		Zijian~Zhang,~\IEEEmembership{Student Member,~IEEE,}
		Linglong~Dai,~\IEEEmembership{Fellow,~IEEE,}
		and~Chan-Byoung Chae,~\IEEEmembership{Fellow,~IEEE}
	}
	\thanks{This work was supported in part by the National Key Research and Development Program of China (Grant No. 2020YFB1807201), in part by the National Natural Science Foundation of China (Grant No. 62031019), and in part by the European Commission through the H2020-MSCA-ITN META WIRELESS Research Project under Grant 956256. This work was also supported by the Korea government (No. 2021-0-00486, 2021-0-02208).  \it{(Corresponding author: Linglong Dai.)}}
	\thanks{Zhongzhichao~Wan, Jieao~Zhu, Zijian~Zhang and Linglong~Dai are with the Department of Electronic Engineering, Tsinghua University as well as Beijing National Research Center for Information Science and Technology (BNRist), Beijing 100084, China (E-mails: \{wzzc20, zja21, zhangzj20\}@mails.tsinghua.edu.cn; daill@tsinghua.edu.cn). }
		\thanks{	Chan-Byoung Chae is with the School of Integrated Technology, Yonsei University, Seoul 03722, Korea (E-mails: cbchae@yonsei.ac.kr).}

}

\maketitle

\begin{abstract}
Traditional channel capacity based on the discrete spatial dimensions mismatches the continuous electromagnetic fields. For the wireless communication system in a limited region, the spatial discretization may results in information loss because the continuous field can not be perfectly recovered from the sampling points. Therefore, electromagnetic information theory based on spatially continuous electromagnetic fields becomes necessary to reveal the fundamental theoretical capacity bound of communication systems. In this paper, we propose analyzing schemes for the performance limit between continuous transceivers. Specifically, we model the communication process between two continuous regions by random fields. Then, for the white noise model, we use Mercer expansion to derive the mutual information between the source and the destination. For the close-form expression, an analytic method is introduced based on autocorrelation functions with rational spectrum. Moreover, the Fredholm determinant is used for the general autocorrelation functions to provide the numerical calculation scheme. Further works extend the white noise model to colored noise and discuss the mutual information under it. Finally, we build an ideal model with infinite-length source and destination which shows a strong correpsondence with the time-domain model in classical information theory. The mutual information and the capacity are derived through the spatial spectral density.
\end{abstract}


%

\section{Introduction}

Wireless communication systems employ electromagnetic fields with three continuous spatial dimensions for information exchange. 
However, the modern multiple-input multiple-output (MIMO) technology, viewed as a discretization of the continuous spatial dimensions, mismatches the continuous nature of electromagnetic fields in real-world communication systems, thus causing its inability to fully explore the spatial information~\cite{sanguinetti2022wavenumber}. 
Therefore, we should restore to the continuous electromagnetic fields to analyze the fundamental performance limit of an arbitrary communication system, motivating the research of the electromagnetic information theory (EIT)~\cite{gruber2008new}. 

For EIT, one asymptotic approach to analyze the fundamental limit is called the spatial bandwidth \cite{bucci1980fast}, based on which the degrees of freedom (DoF) can be theoretically derived. 
The spatial bandwidth of scattered fields under a time-harmonic model was rigorously derived in~\cite{bucci1987spatial}. 
Further works extended the time-harmonic model, which was focused on a single frequency point, to a more general band-limited model, and analyzed the DoF~\cite{franceschetti2015landau}. Unfortunately, these spatial bandwidth-based procedures rely on the assumption that the occupied region of the information destination tends to infinity. 
In contrast, communication systems are often confined in a finite-sized space in practice. In this practical scenario, the spatial bandwidth method becomes inaccurate due to the condition of the Nyquist sampling theorem. 

For the practical communication system in a limited space, several works utilized the orthogonal expansion techniques to derive the DoF for EIT. For example, inspired by the orthogonal frequency-division multiplexing (OFDM), the authors in~\cite{sanguinetti2022wavenumber} designed a set of bases to expand the Green's function, and derived the DoF between two parallel finite-length linear antennas. Moreover, for the scenario where the two finite-length linear antennas bear an intersection angle, a heuristic method was proposed to construct the bases for the derivation of DoF \cite{decarli2021communication}. 
These works provide us with useful methods for analyzing the DoF in EIT. However, they heavily rely on the assumption of deterministic signals, thus being unable to derive the information-theoretic capacity where stochastic properties should first be modeled.

To derive the capacity of the electromagnetic channels, several works introduced basis expansion to split the continuous electromagnetic channel to almost orthogonal channels. The approximation of the basis based on the radiating term of the Green's function was derived~\cite{jensen2008capacity}. The orthogonal bases between a pair of concentric spherical source and destination were shown to be spherical harmonic functions \cite{gruber2008new}. Moreover, the capacity of the system with spherical source and destination in lossy medium was obtained in \cite{jeon2017capacity}. 
Another important approach using the Kolmogorov $\epsilon$-capacity was utilized in \cite{migliore2008electromagnetics} and \cite{migliore2018horse}, which was based on the maximum amount of information transmissible through the channel with an uncertainty level. 

These works already found the best bases for the electromagnetic channels and derived the capacity with finite bases. The capacity can be viewed as the mutual information under a specific field distribution, where the mutual information is maximized. However, an analytic solution and calculation scheme for the mutual information between the transceivers with arbitrary given field distribution has not been obtained in the literature of EIT. Such analysis requires the analytical analyzing scheme based on infinite bases decomposed from the continuous channel. Moreover, the existing works considered the spatial white Gaussian noise field. Since white noise is only a special case of colored noise field and the practical noise field may include non-white component, further works are necessary to explore the mutual information and capacity under colored noise scenario.   

Different from existing works, in this paper, we first build a model considering the non-white component of the noise field. Then, we provide a strict analysis framework of the mutual information and capacity based on the random field theory and operator theory\footnote{Simulation codes are provided to reproduce the results in this paper: \url{http://oa.ee.tsinghua.edu.cn/dailinglong/publications/publications.html}.}. The electromagnetic waves that carry information are modeled as random fields, which follow the statistical approach of Shannon. We introduce the theory of operator analysis to derive the general expression of the mutual information and provide the numerical calculation scheme of it. Specifically, the contributions of this paper are summarized as follows:
\begin{itemize}
	\item{First, we develop a system model of the wireless communication between two continuous regions. We use random fields to capture the statistical characteristics of the signals and noise in the communication system. The mutual information between the source and the destination is defined by a supremum taken over all the testing functions.}
	\item{Then, we consider a simplified model of the communication with finite-length transceivers under white noise field. By exploiting Mercer expansion, we derive the mutual information between the source and the destination. Next, we introduce the analytical scheme of deriving the mutual information for autocorrelation functions with rational spectrum. A special case, which has been adopted in the existing scattering models, is analyzed to show how the analytical scheme works. To generalize the information formulas, we introduce a tool called the Fredholm determinant to obtain a general expression of the mutual information, which enables numerical calculation and convergence analysis. Some discussions about the extendability of the above schemes are also included.}
	\item{Moreover, we adopt the operator analysis schemes to extend the mutual information formula from white noise model to more general cases with colored noise. A discretization scheme is provided for the numerical calculation. Numerical results show the convergence of the calculation scheme and also provide some insights into the performance comparison between continuous and discrete systems.}
	\item{Finally, to illustrate the close relationship between the proposed random field-based theory and the classical time-domain information theory, we build an ideal model with parallel infinite-length linear source and destination. In this model, the mutual information between the source and the destination is represented by the spatial spectral density (SSD) of the electric field and the noise field on the destination. The optimal current density distribution on the source is derived by variational calculus to achieve the maximum mutual information, i.e., the capacity between the source and the destination.} 
\end{itemize}

%
%
%
%
\begin{figure}
	\centering 
	\includegraphics[height= 5.5cm, width=8cm]{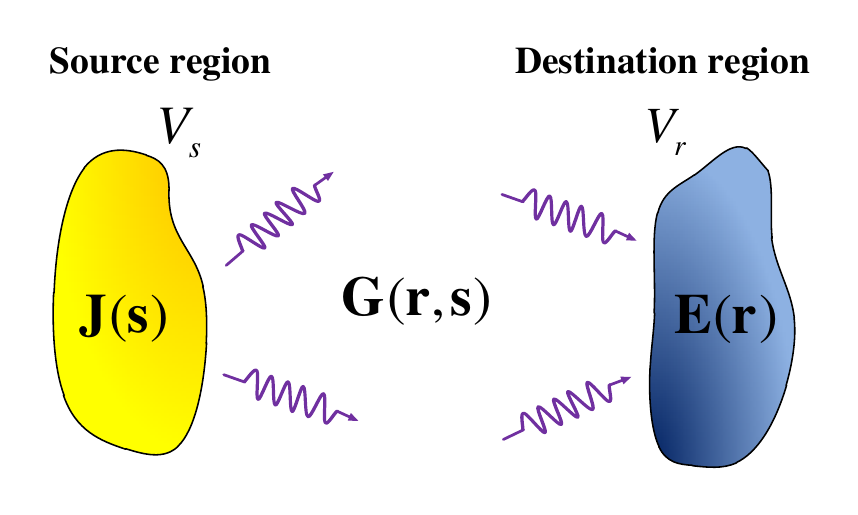} 
	\caption{Electromagnetic model of the communication between two arbitrary continuous regions.} 
	\label{fig:EMmodel}
\end{figure}
\emph{Notation}: bold uppercase characters denote matrices;
bold lowercase characters denote vectors;
the dot $\cdot$ denotes the scalar product of two vectors, or the matrix-vector multiplication. 
${\mathbb E}\left[x\right]$ denotes the mean of random variable $x$; 
$\epsilon_0$ is the permittivity of a vacuum, $\mu_0$ is the permeability of a vacuum, and $c$ is the speed of light in a vacuum; 
$*$ denotes the convolution operation, and $\mathscr{F}[f(x)]$ denotes the Fourier transform of $f(x)$; 
$(f(x))^+$ is equal to $\frac{f(x)+|f(x)|}{2}$; 
$\nabla$ is the nabla operator, and $\nabla \times$ is the curl operator; 
$J_0(x)$ is the Bessel function; 
$K_0(x)$ is the zeroth-order modified Bessel function of the second kind; 
$Y_0(x)$ is the zeroth-order Neumann function;
$\ket{\phi}$ is the quantum mechanical notation of a function $\phi$, where the inner product is denoted by $\bra{\psi}\phi\rangle$;
the matrix element of an operator $T$ under an orthonormal basis $\{\phi_j\}_{j=1}^{+\infty}$ is represented by the Dirac notation $T_{ij}=\bra{i}T\ket{j}$;
$\det(\cdot)$ denotes the matrix determinant or the Fredholm determinant. 

\section{Electromagnetic Wave Communication Model}
\label{sec:EM model}
Maxwell's equations, which consist of four differential equations, describe how electromagnetic fields are generated by currents, charges and the change of fields \cite{griffiths2005introduction}. Here we focus on the two curl equations in Maxwell's equations, which are called the Faraday's law and the Ampère's law, i.e.,
		$\nabla  \times {\bf{E}} =  - \frac{{\partial {\bf{B}}}}{{\partial t}}$ and 
		$\nabla  \times {\bf{H}} = {\bf{J}} + \frac{{\partial {\bf{D}}}}{{\partial t}}$.

These equations are the fundamental physical laws that govern the electromagnetic wave communications. 
To simplify the analysis, we adopt the common assumption that the electromagnetic wave oscillates on a single frequency point, which is the well-known time-harmonic assumption \cite{gruber2008new}. 
In this way, the temporal derivative operator $\partial/\partial t$ can be replaced by $-{\rm j}\omega$ in Maxwell's equations. 
The time-harmonic assumption simplifies the Maxwell's equations into complex-valued partial differential equations involving only spatial derivatives, i.e.,
$\nabla  \times {\bf{E}} = {\rm j}\omega {\bf{B}}$ and 
$\nabla  \times {\bf{H}} = {\bf{J}} - {\rm j}\omega {\bf{D}}$.
From the above equations we can obtain the vector wave equation \cite{dardari2020communicating} as
\begin{equation}
\nabla  \times \nabla  \times {\bf{E}}\left( {\bf{r}} \right) - {\kappa_0 ^2}{\bf{E}}\left( {\bf{r}} \right) = {\rm j}\omega {\mu _0}{\bf{J}}\left( {\bf{r}} \right) = {\rm j}\kappa_0 {Z_0}{\bf{J}}\left( {\bf{r}} \right),
\label{VWE}
\end{equation}
where $\kappa_0  = \omega \sqrt {\mu_0 \varepsilon_0 }$ is the wavenumber and $Z_{0}=\mu_0 c = 120 \pi\,{\rm  [\Omega]}$ is the free-space intrinsic impedance. 

Consider two arbitrary regions $V_{\rm s}$ and $V_{\rm r}$ as the source and the destination for wireless communications. 
The current density at the source is ${\bf J}({\bf s}) : \mathbb{R}^3 \to \mathbb{C}^3$, and the induced electric field at the destination is ${\bf E}({\bf r})$, where ${\bf r} \in \mathbb{R}^3$ is the coordinate of the field observer. 
The received electric field is ${\bf Y}({\bf r})={\bf E}({\bf r})+{\bf N}({\bf r})$, where ${\bf N}({\bf r})$ is the noise field. 
Exploiting the linear nature of \eqref{VWE}, the Green's function ${\bf{G}} ({\bf{r}},{\bf{s}})\in \mathbb{C}^{3\times 3}$ is introduced to solve this equation.
Utilizing the Green's function, the electric field ${\bf{E}}({\bf{r}})$ can be expressed by  
\begin{equation}
	{\bf{E}}({\bf{r}}) = \int_{{V_s}} {\bf{G}} ({\bf{r}},{\bf{s}}){\bf{J}}({\bf{s}}){\rm d}{\bf{s}},\quad {\bf{r}} \in {V_r}.
	\label{VWE_Green}
\end{equation}
The Green's function in unbounded, homogeneous mediums at a fixed frequency point is  \cite{poon2005degrees}
\begin{align}
		{\bf{G}}({\bf{r}},{\bf{s}}) &= \frac{{\rm j}\kappa_0 {Z_0}}{{4\pi }} \left( {{\bf{I}} + \frac{{{\nabla _{\bf{r}}}\nabla _{\bf{r}}^{\rm{H}}}}{{{\kappa_0 ^2}}}} \right) \frac{{{e^{{\rm{j}}\kappa_0 \left\| {{\bf{r}} - {\bf{s}}} \right\|}}}}{{\left\| {{\bf{r}} - {\bf{s}}} \right\|}}   \notag
		\\&= \frac{{\rm j}\kappa_0 {Z_0}}{{4\pi }}\frac{{{e^{{\rm{j}}\kappa_0 \left\| {{\bf{r}} - {\bf{s}}} \right\|}}}}{{\left\| {{\bf{r}} - {\bf{s}}} \right\|}}\Bigg[\left( {{\bf{I}} - {\bf{\hat p}}{{{\bf{\hat p}}}^{\rm{H}}}} \right) \notag \\&~~+ \frac{{\rm j}}{2\pi \left\| {{\bf{r}} - {\bf{s}}} \right\| /\lambda}\left( {\bf I}-3{\bf{\hat p}}{{{\bf{\hat p}}}^{\rm{H}}} \right) \notag \\&~~-\frac{1}{(2\pi\left\| {{\bf{r}} - {\bf{s}}} \right\|/\lambda )^2 }\left( {\bf I}-3{\bf{\hat p}}{{{\bf{\hat p}}}^{\rm{H}}}  \right) \Bigg] [{\rm \Omega}/{\rm m}^2],
		\label{Green}
\end{align}
where ${\bf{\hat p}} = \frac{{\bf{p}}}{{\left\| {\bf{p}} \right\|}}$ and ${\bf{p}} = {\bf{r}} - {\bf{s}}$. Some existing works \cite{sanguinetti2022wavenumber} \cite{decarli2021communication} adopted far-field assumption and only preserved the item $\frac{{\rm j}\kappa_0 {Z_0}}{{4\pi }}\frac{{{e^{{\rm{j}}\kappa_0 \left\| {{\bf{r}} - {\bf{s}}} \right\|}}}}{{\left\| {{\bf{r}} - {\bf{s}}} \right\|}}\left( {{\bf{I}} - {\bf{\hat p}}{{{\bf{\hat p}}}^{\rm{H}}}} \right)$ in (\ref{Green}). Different from these works, we use the whole expression in (\ref{Green}) to make our analysis and numerical calculation valid for both far-field and near-field scenarios.

In this section, we have reviewed the basic model of electromagnetic wave communication. 
However, the information transmission process naturally involves the elimination of uncertainty, which requires probabilistic modeling of the EM fields.
The amount of information transmitted through the EM field can be viewed as the amount of uncertainty eliminated in the instant the field in $V_r$ is observed. 
Thus, in the next section, we will use Gaussian random fields to capture the statistical properties of electromagnetic fields, and then derive the mutual information. 

\section{Random Field Modeling}
\label{sec:spatial analysis}

In this section, we analyze the statistical properties of the signals and noises. We use Gaussian random fields to model the signals and noise of wireless communication systems, which follows the statistical approach of Shannon. Based on the random field modeling, we define the mutual information in a supremum form, which provides foundations for further analysis. 

\subsection{Random field modeling of the signals}
The information-theoretic analysis in this paper is based on stochastic models according to Shannon's approach, which motivates us to model the continuous EM signals by a random field. Each realization of the random field represents a radiating and receiving pattern during one channel use. All the realizations are gathered together to form a probabilistic ensemble, which reflects the statistical characteristics of the wireless communication system, thus leading to mutual information and capacity.
Among all kinds of random fields, the Gaussian random field is of theoretical significance, since its Gaussian distribution is the capacity-achieving distribution of an AWGN channel \cite{shannon1948mathematical}. As a result, we use Gaussian random field to depict the statistical characteristics of both the current density at the source and the electric field at the destination. The Gaussian random field is assumed to be continuous, separable, and fully characterized by its mean and autocorrelation functions \cite{gelfand1959calculation}. The current density ${\bf{J}}({\bf{s}})$ at the source and the induced electric field ${\bf{E}}({\bf{r}})$ at the destination are considered as random fields with autocorrelation functions given by 
\begin{subequations}
	\begin{align}
	&R_{\bf{J}}({\bf{s}},{\bf{s}}') = {\mathbb E}[{\bf{J}}({\bf{s}}){\bf{J}}^{\rm H}({\bf{s}}')]\;[{\rm A}^2/{\rm m}^4], \\
	& R_{\bf{E}}({\bf{r}},{\bf{r}}') = {\mathbb E}[{\bf{E}}({\bf{r}}){\bf{E}}^{\rm H}({\bf{r}}')]\;[{\rm V}^2/{\rm m}^2].
	\end{align}
\end{subequations}
From (\ref{VWE_Green}), we can derive the relation between the autocorrelation function of the current density and that of the electric field as
\begin{align}
R_{\bf{E}}({\bf{r}},{\bf{r}}') &= {\mathbb E}[{\bf{E}}({\bf{r}}){\bf{E}}^{\rm H}({\bf{r}})] \notag
\\&=\int_{V_s}\int_{V_s}{\bf G}({\bf r},{\bf s})R_{\bf{J}}({\bf{s}},{\bf{s}}'){\bf G}^{\rm H}({\bf r},{\bf s}){\rm d}{\bf s}{\rm d}{\bf s}'.
\end{align}

In this model, we assume that the channel is deterministic, and it only consists of line-of-sight components, which means that the uncertainty of the field measurements only comes from the additive noise field. The model of the noise field will be discussed in the following subsection.

\subsection{Random field modeling of the noise}\label{random_field_modeling_of_noise}
In this subsection, we will analyze the model of the noise field. The noise in the communication between a pair of continuous source and destination can be decomposed into two categories: measurement noise and radiation interference. 

For the measurement noise, the authors of \cite{gruber2008new} attributed this kind of noise to the non-ideal factors in communications, including imperfect antenna locations, imprecise field measurements, and numerical errors inevitable during signal processing. 
Since these defects are usually spatially uncorrelated, the noise is then modeled by a white Gaussian random field, which can be characterized by
\begin{equation}
	\begin{aligned}
	{\mathbb E}\left[{\bf E}_{\rm noise}(\bf r){\bf E}_{\rm noise}^{\rm H}(\bf r')  \right] = \frac{n_0}{2}{\bf I}_3\delta({\bf r}-{\bf r'}),
	\end{aligned}
\end{equation}
where $n_0\;[{\rm V}^2\cdot{\rm m}]$ is the single-sided power spectral density in the three-dimensional wavenumber domain. This measurement noise is identical in distribution with the so-called thermal noise, which is widely used in the literature, but it is a wider notion consisting of all the spatially uncorrected undesired signals. The white Gaussian noise (WGN) assumption of this measurement noise helps simplify our analysis. However, the thermal noise model can not depict the interference from other electromagnetic sources which cause the radiation interference.

For the radiation interference, it can be viewed as the superposition of the incident electromagnetic waves which are not generated by the desired source current \cite{sanguinetti2022wavenumber}. 
The simplest analysis is to assume an isotropic incident wave, i.e., the incident power is uniformly distributed among different spatial directions.
Thus, the electromagnetic wave which impinges on the destination from an arbitrary angle can be represented in the spherical coordinates by an elevation angle $\theta \in [0,\pi)$, and an azimuth angle $\phi \in [-\pi,\pi]$. 
It is derived in \cite{sanguinetti2022wavenumber} that, with this isotropic assumption, the noise field can be characterized by 
\begin{subequations}
	\begin{align}
		&{\mathbb E}\left[ {\bf N}({\bf r}+{\bf r'}){\bf N}({\bf r'})^{\rm H} \right]=\sigma^2 \rho({\bf r}){\bf I}_3,\\
	&\rho({\bf r}) = {\rm sinc}\left(\frac{2\left\|{\bf r} \right\|}{\lambda}\right)={\rm sinc}\left(\frac{2r}{\lambda}\right).
	\end{align}
\label{eqn:spatial_correlation_rho}
\end{subequations} 
Note that from \eqref{eqn:spatial_correlation_rho}, the noise field should also be completely isotropic, i.e., the field components in orthogonal directions should not be correlated. 
Unfortunately, this is not the case in reality. 
This misunderstanding occurs because the physical law that electromagnetic waves are transverse waves was not taken into account in \cite{sanguinetti2022wavenumber}, i.e., the direction of the oscillating electric field should be perpendicular to the wave propagation direction. 
Here we will fix this problem and derive a new noise model based on random fields.

The noise field observed at position $\bf r$ can be derived by integrating the waves ${\bf a}\in\mathbb{C}^3$ incoming $\bf r$ from all spatial directions, which can be represented by
\begin{equation}
	{\bf N}({\bf r}) = \int_{-\pi}^{\pi}\int_0^{\pi}{\bf a}(\theta,\varphi)e^{{\rm j}\boldsymbol{\kappa} \cdot {\bf r}} {\rm d} \theta {\rm d}\varphi,
\end{equation} 
where the wavevector is expressed in spherical coordinates as 
\begin{equation}
	\boldsymbol{\kappa} = \frac{2\pi}{\lambda}\left[{\rm cos}\varphi {\rm sin}\theta,{\rm sin}\varphi {\rm sin}\theta, {\rm cos} \theta \right]\in\mathbb{R}^3.
\end{equation}

Let $\kappa = \left\| \boldsymbol{\kappa} \right\|$, and the normalized propagation direction vector $\boldsymbol{\hat{\kappa}} = \boldsymbol{\kappa}/\kappa$, then we have ${\bf a}^{\rm H}(\theta,\varphi)\boldsymbol{\hat{\kappa}} = 0$, i.e., the incident electric field oscillates in the plane perpendicular to the wavevector $\bf{k}$. By further assuming an uncorrelated random incident phase of the two distinct polarizations, the correlation matrix of the random vector ${\bf a}(\theta,\varphi)$ can be represented by constant multiples of ${\bf I} - \boldsymbol{\hat{\kappa}}\boldsymbol{\hat{\kappa}}^{\rm T}$. Thus, we can obtain the angular autocorrelation of ${\bf a}(\theta,\varphi)$ to be 
\begin{align}
	{\mathbb E}\left[{\bf a}(\theta,\varphi){\bf a}^{\rm H}(\theta',\varphi')  \right] &= \sigma^2 f(\theta, \varphi) (\bf{I}-\boldsymbol{\hat{\kappa}}\boldsymbol{\hat{\kappa}}^{\rm T}) \notag \\&~~~~\delta(\theta-\theta')\delta(\varphi-\varphi'),
	\label{eqn:planar_wave_autocorr}
\end{align} 
where the dimensionless density $f(\theta, \varphi)$ equals $\frac{{\rm sin}\theta}{4\pi} $ because of the isotropic propagation assumption \cite{sanguinetti2022wavenumber}, and $\sigma^2\,{\rm [V^2/m^2]}$ is the average second-order moment of the noise field.
With \eqref{eqn:planar_wave_autocorr}, the autocorrelation function of the noise field can then be derived by integrating incident waves from all the spatial directions over the unit spherical shell:
\begin{align}
		&~~~~{\mathbb E}\left[{\bf N}(\bf r +\bf r'){\bf N}^{\rm H}(\bf r')  \right] \notag \\&= \int_{-\pi}^{\pi}\int_0^{\pi}\int_{-\pi}^{\pi}\int_0^{\pi} {\mathbb E} \left[ {\bf a}(\theta,\varphi){\bf a}^{\rm H}(\theta',\varphi') \right]  e^{{\rm j}\boldsymbol{\kappa}\cdot \left({\bf r}+{\bf r'}\right)} \notag \\&~~~~e^{-{\rm j}\boldsymbol{\kappa'}\cdot {\bf r'}}{\rm d} \theta {\rm d}\varphi{\rm d} \theta' {\rm d}\varphi' \notag \\
		& = \int_{-\pi}^{\pi}\int_0^{\pi}\sigma^2 f(\theta, \varphi) (\bf{I}-\boldsymbol{\hat{\kappa}}\boldsymbol{\hat{\kappa}}^{\rm T})e^{{\rm j}\boldsymbol{\kappa}\cdot {\bf r}}{\rm d} \theta {\rm d}\varphi \notag
		\\&= \frac{\sigma^2}{4\pi}\underset{S_3}{\iint}({\bf{I}-\boldsymbol{\hat{\kappa}}}\boldsymbol{\hat{\kappa}}^{\rm T})e^{{\rm j}\boldsymbol{\kappa}\cdot {\bf r}}{\rm d} S,
	\label{eqn:noise_autocorr_primitive}
\end{align} 
where ${\rm d}S$ denotes the area element on the unit spherical shell. To solve the definite integral in \eqref{eqn:noise_autocorr_primitive}, we define two auxiliary functions $f_1(\beta)$ and $f_2(\beta)$ as
\begin{subequations}
	\begin{align}
		&f_1(\beta) = \int_{-1}^{1}e^{{\rm j}\beta x}{\rm d}x = 2\frac{{\rm sin}\beta}{\beta},
		\\&f_2(\beta) = \int_{-1}^{1}x^2e^{{\rm j}\beta x}{\rm d}x = 2\left(\frac{{\rm sin}\beta}{\beta}+\frac{2{\rm cos}\beta}{\beta^2}-\frac{2{\rm sin}\beta}{\beta^3}\right).
	\end{align}
\end{subequations} 
Then, we have
\begin{align}
		\frac{\sigma^2}{4\pi}\underset{S_3}{\iint}\bf{I}e^{{\rm j}\boldsymbol{\kappa}\cdot {\bf r}}{\rm d}S &= \frac{\sigma^2}{4\pi}\int_{-\pi}^{\pi}\int_0^{\pi}{\bf{I}}e^{{\rm j}\boldsymbol{\kappa}\cdot {\bf r}}{\rm sin}\theta{\rm d} \theta {\rm d}\varphi
		\notag \\&=\frac{\sigma^2}{4\pi}\int_{-\pi}^{\pi}\int_0^{\pi}{\bf I}e^{{\rm j}{\kappa}r{\rm cos}\theta}{\rm sin}\theta{\rm d} \theta {\rm d}\varphi \notag
		 \\&= \frac{\sigma^2}{2}{\bf{I}} f_1(\kappa r).
\end{align} 
For the term $\frac{\sigma^2}{4\pi}\iint_{S_3}\boldsymbol{\hat{\kappa}}\boldsymbol{\hat{\kappa}}^{\rm T}e^{{\rm j}\boldsymbol{\kappa}\cdot {\bf r}}{\rm d}S$, we consider its component along $\bf{\hat{r}}$ and perpendicular to $\bf{\hat{r}}$ sequentially to derive its analytical representation. For the component along $\bf{\hat{r}}$, we have
\begin{align}
	\frac{\sigma^2}{4\pi}\underset{S_3}{\iint}\boldsymbol{\hat{\kappa}}\boldsymbol{\hat{\kappa}}^{\rm T}e^{{\rm j}\boldsymbol{\kappa}\cdot{\bf r}}{\rm d}S \cdot {\bf{\hat{r}}} &= \frac{\sigma^2}{4\pi}\underset{S_3}{\iint}{\boldsymbol{\hat{\kappa}}}(\boldsymbol{\hat{\kappa}}^{\rm T}{\bf{\hat{r}}})e^{{\rm j}\boldsymbol{\kappa}\cdot {\bf r}}{\rm d}S  \notag
	\\&\overset{(a)}{=} \frac{\sigma^2}{2} \int_0^{\pi}{\rm sin}\theta {\rm cos}^2\theta e^{{\rm j}{ \kappa}{r}{\rm cos}\theta} {\rm d} \theta {\bf{\hat{r}}} \notag
	 \\&= \frac{\sigma^2}{2} f_2(\kappa r) {\bf{\hat{r}}},
\end{align} 
where $(a)$ comes from the symmetric property of the integral along the $\bf{\hat{r}}$ axis. In order to solve the components which are perpendicular to $\bf{\hat{r}}$, we assume that ${\bf{\hat{t}}}_1$ and ${\bf{\hat{t}}}_2$ are two orthogonal directions on the plane perpendicular to $\bf{\hat{r}}$, where the direction cosines of unit vector ${\bf \hat{\kappa}}$ under the newly defined coordinate system $({\bf{\hat{t}}}_1, {\bf{\hat{t}}}_2,{\bf{\hat{r}}})$ are $({\rm sin}\theta {\rm cos}\varphi, {\rm sin}\theta {\rm sin}\varphi, \cos\theta)$. Then, we can derive
\begin{align} 
	&~~~~{\bf{\hat{t}}}_1 \cdot\frac{\sigma^2}{4\pi}\underset{S_3}{\iint}{\boldsymbol{\hat{\kappa}}}{\boldsymbol{\hat{\kappa}}}^{\rm T}e^{{\rm j}\boldsymbol{\kappa}\cdot {\bf r}}{\rm d}S \cdot {\bf{\hat{t}}}_1 \notag \\&= \frac{\sigma^2}{4\pi}\underset{S_3}{\iint} ({\rm sin}\theta {\rm cos}\varphi)^2 e^{{\rm j}\boldsymbol{\kappa}\cdot {\bf r}}{\rm d}S
	\notag \\&= \frac{\sigma^2}{4\pi}\int_{-\pi}^{\pi}\int_0^{\pi} {\rm sin}^3 \theta {\rm cos}^2 \varphi e^{{\rm j}{ \kappa}{ r}{\rm cos}\theta}{\rm d}\theta {\rm d} \varphi \notag
	\\& = \frac{\sigma^2}{4} \int_0^{\pi} {\rm sin}^3 \theta e^{{\rm j}{ \kappa}{ r}{\rm cos}\theta}{\rm d}\theta = \frac{\sigma^2}{4}(f_1(\kappa r)-f_2(\kappa r)),
\end{align} 
and
\begin{equation}
	\begin{aligned}
	{\bf{\hat{t}}}_2 \cdot\frac{\sigma^2}{4\pi}\underset{S_3}{\iint}{\boldsymbol{\hat{\kappa}}\boldsymbol{\hat{\kappa}}^{\rm T}}e^{{\rm j}\boldsymbol{\kappa}\cdot {\bf r}}{\rm d}S \cdot {\bf{\hat{t}}}_2 &= \frac{\sigma^2}{4\pi}\underset{S_3}{\iint} ({\rm sin}\theta {\rm sin}\varphi)^2 e^{{\rm j}\boldsymbol{\kappa}\cdot {\bf r}}{\rm d}S\\&= \frac{\sigma^2}{4}(f_1(\kappa r)-f_2(\kappa r)).
	\end{aligned} 
\end{equation} 

Similarly, we can prove that the components ${\bf \hat{t}}_i\cdot\iint_{S_3}{\boldsymbol{\hat{\kappa}}}{\boldsymbol{\hat{\kappa}}}^{\rm T}e^{{\rm j}{\bf k}\cdot {\bf r}}{\rm d}S\cdot{\bf \hat{t}}_j=0$ for all $i\neq j, i,j\in\{1,2\}$. Combining all the above results into a 3-by-3 autocorrelation matrix, we have
\begin{align}
&~~~~{\mathbb E}\left[{\bf N}(\bf r +\bf r'){\bf N}^{\rm H}(\bf r')  \right] \notag \\&= \frac{\sigma^2}{2} f_1(\kappa r){\bf{I}}-\frac{\sigma^2}{2} f_2(\kappa r) {\bf{\hat{r}}\bf{\hat{r}}}^{\rm T} \notag \\&~~~~-\frac{\sigma^2}{4}(f_1(\kappa r)-f_2(\kappa r))({\bf{I}}-{\bf{\hat{r}}}{\bf{\hat{r}}}^{\rm T})
\notag \\&=\frac{\sigma^2}{4}(f_1(\kappa r)+f_2(\kappa r)){\bf{I}}+\frac{\sigma^2}{4}(f_1(\kappa r)-3f_2(\kappa r)){\bf{\hat{r}}}{\bf{\hat{r}}^{\rm T}}.
\label{equ:noise_model_radiated}
\end{align} 
Specifically, the noise field correlation measured along the polarization direction ${\bf \hat{r}}$ is given by 
\begin{equation}
	\begin{aligned}
		{\mathbb E}\left[ N(r+r')N(r') \right] &= {\bf{\hat{r}}^{\rm H}} {\bf R}_{\bf N} {\bf{\hat{r}}} =\frac{\sigma^2}{2} f_1(\kappa r)-\frac{\sigma^2}{2} f_2(\kappa r) \\&= 2 \sigma^2 \left(\frac{{\rm sin }(\kappa r)}{(\kappa r)^3}-\frac{{\rm cos}(\kappa r)}{(\kappa r)^2}\right).
	\end{aligned}
	\label{eqn:parallel_direction}
\end{equation}
The result can also be applied to a uniform linear antenna array placed on the $x$-axis, where the polarization orientation is parallel to the $y$-axis. With this assumption, the noise field correlation is given by 
\begin{equation}
	{\mathbb E}\left[N(r+r')N(r')\right] = \frac{\sigma^2}{4}(f_1(\kappa r)+f_2(\kappa r)).
\end{equation}
This equation reveals that different from our previous understanding of the noise that its autocorrelation is of the form ${\rm sinc}(\cdot)$. The precise noise autocorrelation contains more high-spatial-frequency components, which come from the term $f_2(\kappa r)$. 

\begin{remark}
	Equation \eqref{equ:noise_model_radiated} fully depicts the autocorrelation of the noise field of three polarization orientations under the assumption of isotropic incidence. This form of autocorrelation is also employed to describe the channel autocorrelation \cite[eqn (19)]{shafi2006polarized}, which is reasonable because the noise discussed here can be considered as unwanted signals received through a random channel. 
\end{remark}
\begin{remark}
	This noise model is based on the isotropic scattering assumption, which can be easily extended to the non-isotropic case. If we assume that the radiation interference is not uniform in spatial angle but concentrated near a certain angle, we can use von Mises-Fisher (vMF) distribution \cite{abdi2002space, gatto2007generalized} to depict such angular concentration. A similar idea was proposed in \cite{pizzo2022spatial} to analyze the angular selectivity of the random channel modeling but the analytic solution was not given. For the 3-dimensional von Mises-Fisher distribution, which is a close approximation to the spherical analogue of the Gaussian distribution, the probability density function is given by 
	\begin{equation}
		\begin{aligned}
			f(\boldsymbol{x}|\boldsymbol{\mu}) = C(\|\boldsymbol{\mu}\|)e^{\hat{\boldsymbol{x}} \cdot \boldsymbol{\mu}},
		\end{aligned}
	\end{equation} 
	where $C(\|\boldsymbol{\mu}\|) = \frac{\|\boldsymbol{\mu}\|}{2\pi (e^{\|\boldsymbol{\mu}\|}-e^{-\|\boldsymbol{\mu}\|})}$ is the normalization constant. Instead of (\ref{eqn:noise_autocorr_primitive}) we have
	\begin{equation}
		\begin{aligned}
			{\mathbb E}\left[{\bf N}(\bf r +\bf r'){\bf N}^{\rm H}(\bf r')  \right] 
			= \frac{\sigma^2}{4\pi}\underset{S_3}{\iint}({\bf{I}-\boldsymbol{\hat{\kappa}}}\boldsymbol{\hat{\kappa}}^{\rm T})e^{{\rm j}\boldsymbol{\kappa}\cdot {\bf r}}e^{\hat{\boldsymbol{\kappa}} \cdot \boldsymbol{\mu}}{\rm d} S.
		\end{aligned}
	\end{equation}
	
	Then, by replacing ${\bf r}\in\mathbb{R}^3$ in (\ref{eqn:noise_autocorr_primitive}) by ${\bf r} = {\bf r}_R - {\rm j}{\boldsymbol \mu}/\kappa \in\mathbb{C}^3$. The imaginary part ${\rm j}{\boldsymbol \mu}/k$ can be combined into the spatial harmonic factor $e^{{\rm j}\boldsymbol{\kappa} \cdot {\bf r}}$ of the planar wave, in order to describe a vMF distribution of the incident angles $\theta$ and $\varphi$. Equation \eqref{equ:noise_model_radiated} can then be extended to the non-isotropic case accordingly by analytic continuation \cite{ahlfors1953complex} techniques, leading to
	\begin{equation}
		\begin{aligned}
			{\mathbb E}\left[{\bf N}(\bf r +\bf r'){\bf N}^{\rm H}(\bf r')  \right] &=\frac{\sigma^2}{4}(f_1(\kappa r_1)+f_2(\kappa r_1)){\bf{I}}\\&~~~~+\frac{\sigma^2}{4}(f_1(\kappa r_1)-3f_2(\kappa r_1)){\bf{\hat{r}_1}}{\bf{\hat{r}_1}^{\rm T}},
		\end{aligned}
	\end{equation} 	 
	where $r_1 = \sqrt{(\boldsymbol{r}-{\rm j}\boldsymbol{\mu}/\kappa)^{\rm T}(\boldsymbol{r}-{\rm j}\boldsymbol{\mu}/\kappa)}$ and ${\bf{\hat{r}_1}} = \frac{\boldsymbol{r}-{\rm j}\boldsymbol{\mu}/\kappa}{r_1}$.
\end{remark}

\subsection{Mutual information based on random field modeling}
The mutual information was originally defined by Shannon in his seminal paper \cite{shannon1948mathematical} to be the amount of uncertainty reduced by observations of the channel outputs. The channel usually admits discrete random input symbols, and gives out discrete output symbols. 
However, in our study on electromagnetic information theory, the channel output ${\bf E}({\bf r})$ is a spatially continuous electromagnetic field, to which we assign uncertainty by modeling it as a Gaussian random field. 
Similar to the standard definition of mutual information, upon obtaining noisy measurements ${\bf Y}({\bf r})$ of this field, we can also evaluate the amount of information received through the uncertainty reduction mechanism, but the field measurements should be well-defined to avoid the ``continuum'' difficulties encountered when analyzing a spatially continuous field. 

As introduced in the above subsection, the noise is modeled as a Gaussian random field with autocorrelation function $R_{\bf{N}}({\bf{r}},{\bf{r}}')$ \cite{sanguinetti2022wavenumber} and the autocorrelation function of the noisy electric field is denoted by $R_{\bf Y}({\bf r},{\bf r'})$. 
Following the definition in \cite{gelfand1959calculation}, we use testing functions to define the mutual information between the random fields ${\bf J}$ and ${\bf Y}$ by
\begin{equation}
	I({\bf J};{\bf Y}) = {\rm sup}\{I({\bf J}(\phi_1,\cdots,\phi_m),{\bf Y}(\psi_1,\cdots,\psi_n))\},
	\label{equ:mutual_info_definition} 
\end{equation}
where ${\bf J}(\phi_1,\dots,\phi_m) = \{ \langle \phi_1|{\bf J} \rangle ,\cdots, \langle \phi_m|{\bf J} \rangle\}$ and ${\bf Y}(\psi_1,\dots,\psi_n) = \{\langle \psi_1|{\bf Y}\rangle,\cdots,\langle \psi_n|{\bf Y}\rangle\}$ are random vectors, and the inner product $\langle\phi|{\bf J}\rangle$ equals to $\int_{V_s}\phi^*(\bf s){\bf J}({\bf s}){\rm d}{\bf s}$. The supremum is taken over all the possible integers $m$, $n$, and testing functions $\phi,\psi\in\Phi$, where $\Phi$ is the set of all the smooth functions that vanish outside $V_s$.

According to \cite[{\bf Theorem 1.3}]{gelfand1959calculation}, we can select a sequence of $\phi_i^{k}$ which converges to $\delta({\bf s}-{\bf s}_i)$ when $k \rightarrow \infty$. Then, the mutual information $I({\bf J},{\bf Y})$ can be calculated by the mutual information between two groups of points, expressed as ${\rm sup}\{I({\bf J}({\bf s}_1,\cdots,{\bf s}_m),{\bf Y}({\bf r}_1,\cdots,{\bf r}_n))\}$. This means that we can use the random variables defined on the spatial sampling points $\{{\bf s}_1,\cdots, {\bf s}_m \}$ and $\{{\bf r}_1,\cdots, {\bf r}_n \}$ to approximate the mutual information between the two random continuous electromagnetic fields. It is worth noting that this claim only shows the theoretical existence of the sampling points, which does not involve the way to choose them. In this paper, we will show a discretization scheme of the continuous field and the convergence is provided according to numerical analysis.

In this section, the mutual information between continuous regions is defined in a supremum form using random fields. Then, in the next section, we will simplify the model with continuous regions to obtain an analytical solution to the mutual information.

\section{Mutual Information with Finite-Length Transceivers under White Noise}
\label{sec: upperbound}
In this section, to obtain more insightful results, we consider the simplified case where a pair of parallel finite-length linear source and destination are employed as transceivers. White noise model is used in this section for simplicity and non-white noise model will be discussed in the next section for generality. Since the simplified model here is similar to the classical stochastic process in the one-dimensional time domain, most of the existing analyzing schemes and tools can be transfered to the analysis in this simplified scenario. To make the schemes more general, we also show how the adopted tools and results from stochastic process can be extended to random field with at least two dimensions. Specifically, to evaluate the mutual information, we apply Mercer expansion \cite{mercer1909xvi} as a basic mathematical tool. 
Furthermore, we show how to derive analytical solutions of the mutual information and provide an example. Finally, we introduce Fredholm determinant to provide a general expression of the mutual information. The expression based on Fredholm determinant has numerical evaluation scheme whose convergence is guaranteed in the literature.

\begin{figure}
	\centering 
	\includegraphics[width=7cm]{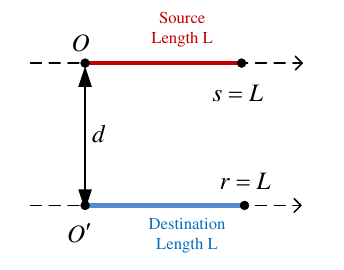}
	\caption{Analysis of mutual information between a source and destination, both of finite length. }
	\label{fig:two_finite}
\end{figure}

\subsection{Mercer expansion of a Gaussian random field}

Assume that the source and the destination are two linear parallel length-$L$ continuous antenna arrays placed along the $z$-axis and 
separated by a distance of $d$. 
Without loss of generality, we only care about the fields along the $z$-axis, i.e., the vector field ${\bf E}({\bf r})$ is only observed at the $z$-direction.
In the following analysis, the scalars $J(s)$ and $E(r)$ represent ${\bf J}(\bf s)\cdot \hat{e}_z$ and ${\bf E}(\bf r)\cdot \hat{e}_z$, respectively.
The second moments (autocorrelation) of $J(s)$ are denoted by $R_J(s,s'), s,s'\in [-L/2,+L/2]$. 

Through this approach, we have built a simplified model which is highly similar to the classical model of random processes in information theory. Then it will be convenient for us to use the tools in classical information theory to solve and analyze the performance of EIT systems, with the transformation from the time-frequency domain to the space-wavenumber domain. Furthermore, we show that the tools for random processes can be extended to multi-dimensional random fields, which can solve other scenarios like rectangular transceivers in EIT. 

The relationship between $J(s)$ and $E(r)$ can be described using the element in the upper left corner of the matrix $\bf G$ in \eqref{Green}, i.e., $G_{1,1}$, which can be derived as
\begin{equation}
\begin{aligned}
\label{equ:Green_g}
{g}{(r,s)} =& \frac{{{\rm j} {Z_0} {e^{{\rm j}2\pi \sqrt {{x^2} + {d^2}} /\lambda }}}}{{2\lambda \sqrt {{x^2} + {d^2}} }}\Big[ \frac{\rm j}{{2\pi \sqrt {{x^2} + {d^2}} /\lambda }}\frac{{{d^2} - 2{x^2}}}{{{x^2} + {d^2}}} 
\\&+\frac{{{d^2}}}{{{x^2} + {d^2}}}
- \frac{1}{{{{(2\pi /\lambda )}^2}({x^2} + {d^2})}}\frac{{{d^2} - 2{x^2}}}{{{x^2} + {d^2}}}\Big],
\end{aligned}
\end{equation}  
where $x = r-s$ and $d$ is the distance between the parallel source and destination. Therefore, we have
	$E(r) = \int_0^L g(r,s)J(s){\rm d}s$.

Since the noiseless received field is uniquely determined by the source and the deterministic channel, the autocorrelation function of the receiving electric field is expressed by the source autocorrelation $R_J(s,s')$ and the Green's function $g(r,s)$, written as
\begin{equation}
	\begin{aligned}
	R_E(r,r') &= \int_{0}^{L} \int_{0}^{L} g(r,s)R_J(s,s') g^{*}(r',s'){\rm d}s{\rm d}s'.
	\end{aligned}
\end{equation}

In order to evaluate $I({\bf J};{\bf Y})$, according to \eqref{equ:mutual_info_definition}, a set of supremum-achieving testing functions $\{\phi_k\}_{k=1}^{+\infty}$ are used to convert the random field into a sequence of pairs of random variables $(\langle \phi_k|{\bf J} \rangle$, $\langle \phi_k|{\bf Y} \rangle)_{k=1}^{+\infty}$ for further information-theoretic analysis. 
Fortunately, the Mercer expansion provides us with a powerful tool to decompose random fields into random variables, as long as the autocorrelation function is given. The supremum-achieving of this expansion is based on the \cite{gelfand1959calculation}, as long as $I({\bf J};{\bf Y})<\infty$. 
According to Mercer's theorem \cite{mercer1909xvi}, the Mercer expansion of the electric field is
\begin{equation}
\begin{aligned}
R_E(r,r') = \sum_{k=1}^{+\infty}\lambda_k\phi_k(r)\phi_k^{*}(r'),
\label{equ:mercer}
\end{aligned}
\end{equation}
where $\phi_k(r)$ are the solution functions to the integral eigen problem
\begin{equation}
\begin{aligned}
\lambda_k \phi_k(r') = \int_{0}^{L}R_E(r,r')\phi_k(r){\rm d}r; k>0,k \in {\mathbb N}.
\label{equ:integral_function}
\end{aligned}
\end{equation}
The eigenfunctions are orthonormal, satisfying the following equation
\begin{equation}
\begin{aligned}
\int_{0}^{L}\phi_{k_1}(r)\phi_{k_2}^{*}(r){\rm d}r = {\mathbbm 1}_{k_1=k_2},
\end{aligned}
\label{equ:orthogonality}
\end{equation}
where the eigenvalues $\{\lambda_k\}_{k=1}^{+\infty}$ are arranged in descending order. 
The field $E(r)$ can be decomposed by $E(r) = \sum_{k=1}^{\infty}\xi_k \phi_k(r)$, where $\mathbb{E}[\xi_{k_i}\xi_{k_j}^{*}] = \lambda_{k_i} {\mathbbm 1}_{i=j}$.
In fact, other orthogonal bases can be chosen instead of $\phi_k(r)$, but only Mercer expansion can guarantee that $\xi_{k_i}$ and $\xi_{k_j}$ are un-correlated, which makes the mutual information of the decomposed subchannels additive.

Similarly, for the noise field, the Mercer expansion also exists and we denote the eigenfunctions by $\phi_{k}'(r)$ and the eigenvalues by $\lambda_{k}'$.
Now we further assume that the noise field is white Gaussian\footnote{In the next section, we will deal with the much more complicated case where the noise is Gaussian but not necessarily white. } with the noise power spectral density $n_0/(2\sqrt{2\pi})$. 
This assumption greatly simplifies the analysis, because the Gaussian white noise has a simple autocorrelation function $R_N(r,r')=\frac{n_0}{2}\delta(r-r')$, which means that the integral equation
$\frac{n_0}{2} \phi_k(r') = \int_{0}^{L}R_N(r,r')\phi_k(r){\rm d}r; k>0,k \in {\mathbb N}$
holds for arbitrary integrable function $\phi_k(r)$. Therefore, the received electric field and the noise\footnote{The Gaussian white noise field is no longer a ``proper'' random field \cite{neeser1993proper}. It cannot be specified by the finite-dimensional distribution function. Instead, it can only be defined by its projections onto a complete orthonormal set of basis functions. } can be expanded on the same orthogonal bases $\phi_k(r)$, and the mutual information between the electric field $E$ and the received field $Y$ can be expressed by the eigenvalues of \eqref{equ:integral_function}
\begin{equation}
\begin{aligned}
I({E}; {Y})=\sum_{k=1}^{+\infty}{\rm log}\left(1+\frac{\lambda_k}{n_0/2}\right).
\end{aligned}
\label{mercer_capacity}
\end{equation}
The rigorous proof is provided by \cite[{\bf Theorem 1}]{zhujieao}. 

\begin{remark}
	For simplicity, we focus on the linear transceivers in the paper, so as to use the tools and conclusions in the classical theory of stochastic process. However, the analyzing schemes here are not limited to the scenario with linear transceivers, but can be extended to other scenarios like two-dimensional surfaces. The conclusions in stochastic processes should also be extended to those in random fields. For example, if the source is a rectangular surface with current density $J({\bf s}) = J(s_x,s_y)$ and the destination receives electric field $E({\bf r}) = E(r_x,r_y)$ in a rectangular surface. We can also derive the autocorrelation function $R_E({\bf r},{\bf r}')$ of the electric field. Since the rectangular surface is a compact separable metric space if we define the area as the measure on the space, we can perform Mercer expansion on random field $E(r_x,r_y)$ according to \cite{mercer1909xvi}. The expansion is similar to (\ref{equ:mercer}) as $R_E({\bf r},{\bf r}') = \sum_k \lambda_k \phi_k({\bf r})\phi_k({\bf r}')$. Therefore, mutual information expression and the corresponding derivation can also be done under the scenario with two-dimensional transceivers. For three-dimensional transceivers, the extension is similar to the two-dimensional cases. Furthermore, if we want to consider the vector-valued fields as ${\bf J}(\bf s)$ and ${\bf E}(\bf r)$, we can decompose ${\bf E}(\bf r)$ according to the extension of the Mercer theorem \cite{de2013extension}. 
\end{remark}

From (\ref{mercer_capacity}) we have derived the mutual information between the induced electric field and the received field. One interesting phenomenon is that although $J$ and $E$ may not obey a one-to-one correspondence, the equation that $I(E;Y) = I(J;Y)$ still exists. The reason that $J$ and $E$ may not be one-to-one correspondent is when we do Fourier transform on both sides, we have $\mathcal{F}[E] = \mathcal{F}[g]\mathcal{F}[J]$. Since $\mathcal{F}[g]$ has two zeros on $\kappa = \pm \kappa_0$, the corresponding two points in $\mathcal{F}[J]$ can not be solved. Physically this means that the electromagnetic wave with $\kappa = \pm \kappa_0$ can only transmit along the line of the source and can not be received by the destination. 

Next we will show why $I(J;Y) = I(E;Y)$. This is from the equality that $I(J;Y)+I(Y;E|J) = I(Y;E)+I(Y;J|E)$.
	On the one hand, we can conclude that $I({J}; {Y}) \leqslant I({E}; {Y})$ by the data processing inequality \cite{polyanskiy2014lecture}, because $J \rightarrow E \rightarrow Y$ forms a Markov chain and $I(Y;J|E)=0$. On the other hand, since $E(r) = \int_{0}^{L} {g} (r,s)J(s){\rm d}{s}$ is a deterministic function, we have $I(Y;E|J) = 0$, which leads to
	 $I({J}; {Y}) \geqslant I({E}; {Y})$. Therefore, we have $I(J;Y) = I(E;Y)$. 

\begin{remark}
	Our analysis here from the mutual information perspective is compatible with existing works about the capacity. The existing works decompose the current density and electric field on a set of bases, which is similar to the SVD decomposition on ${\bf H}{\bf H}^\dagger$, where ${\bf H}$ is the channel matrix in discrete MIMO systems \cite{miller2019waves}. Our work decomposes the current density and electric field according to their autocorrelation function, which finds the bases for the field with given autocorrelation function. This scheme is similar to the decomposition of the covariance matrix to obtain the mutual information between random vectors, since 
        $I(X;X+N) = {\rm logdet} \frac{K_X+K_N}{K_N} = \sum_{i=1}^m (1+\frac{\lambda_m}{n_0/2})$,
    where $\lambda_k$ is the eigenvalue of $K_X$ and $K_N = (n_0/2){\bf I}$.
	
	It is obvious that when fixing the field distribution which maximizes the mutual information, the capacity between continuous transceivers can be obtained. This conclusion is an extension from the discrete matrices to continuous operators. For discrete matrices we have $C = \underset{K_J}{\rm max}{\rm logdet}\frac{HK_JH^{\rm H}+K_N}{K_N}$ with respect to ${\rm Tr}(K_J) \leqslant P$. The optimal solution is obtained by using singular value decomposition and Karush–Kuhn–Tucker (KKT) theorem. For operators, Mercer expansion is utilized instead of the singular value decomposition. Through this approach, an operator can be viewed as an infinitely large matrix \cite{gruber2008new}, and the optimization problem can be solved by extended KKT theorem \cite{bachir2021finitely} for countably infinite variables. The result is in the form of water-filling method.

\end{remark}

	The added values of the mutual information obtained in the paper can be listed as follows: 1) we can derive the mutual information under arbitrary given field distribution. If the field distribution can only be chosen from several given modes instead of the best one when capacity is achieved, the mutual information analyzing scheme can help to find the performance bound. 2) closed-form expression and numerical calculation scheme whose convergent rate is guaranteed can be obtained from the scheme we adopted, which will be introduced in the following subsections. 3) mutual information under colored noise field can be obtained and the corresponding numerical calculation scheme is proposed, which will be presented in {\bf Section. \ref{sec:colored noise field}}.

\subsection{Analytic expression based on autocorrelation function with rational spectrum}

From the above subsection, we have 
\begin{equation}
	I({J}; {Y}) = I({E}; {Y}) = \sum_{k=1}^{+\infty}{\rm log}\left(1+\frac{\lambda_k}{n_0/2}\right).
	\label{equ:I_infinite_sum}
\end{equation}
This formula is hard to evaluate, since there are infinite number of $\lambda_K$ to calculate.
In this part, we will first show how to derive an analytic expression without calculating $\lambda_k$ in \cite{gelfand1959calculation}. Then, we will propose a specific scenario and derive the corresponding analytic expression of mutual information. 

For $R_E(r,r') = \int_{-\infty}^{\infty}e^{{\rm j}(r-r')x}\frac{|Q({\rm j}x)|^2}{|P({\rm j}x)|^2}{\rm d}x$, where $Q({\rm j}x) = \sum_{i=0}^n b_i({\rm j}x)^i$ and $P({\rm j}x) = \sum_{i=0}^n a_i({\rm j}x)^i$, the eigen-problem can be expressed by $\lambda \phi(r) = \int_0^L \int_{-\infty}^{\infty}e^{{\rm j}(r-r')x}\frac{|Q({\rm j}x)|^2}{|P({\rm j}x)|^2}{\rm d}x \phi(r'){\rm d}r'.$ By performing $P(\frac{\rm d}{{\rm d}r})P(-\frac{\rm d}{{\rm d}r})$ on the integral equation, we have 
\begin{equation}
	\begin{aligned}
		P(\frac{\rm d}{{\rm d}r})P(-\frac{\rm d}{{\rm d}r})\phi(r) = \frac{2\pi}{\lambda}Q(\frac{\rm d}{{\rm d}r})Q(-\frac{\rm d}{{\rm d}r})\phi(r).
        \label{equ:transformed_differential_equation}
	\end{aligned}
\end{equation} 
By performing $\frac{{\rm d}^k}{{\rm d}r^k}P(\frac{\rm d}{{\rm d}r})~~(k = 0,\cdots, n-1)$ on the integral equation, we have boundary conditions
\begin{equation}
	\begin{aligned}
		\frac{{\rm d}^k}{{\rm d}r^k}P(\frac{\rm d}{{\rm d}r})\phi(r) &= \frac{1}{\lambda}\int_0^L \int_{-\infty}^{\infty}e^{{\rm j}(r-r')x}\frac{|Q({\rm j}x)|^2({\rm j}x)^k}{P(-{\rm j}x)}{\rm d}x \\&~~~~\phi(r'){\rm d}r',
        \label{equ:boundary}
	\end{aligned}
\end{equation} 
where $r$ is set on $0$ and $L$ to obtain the boundary conditions. 

By solving (\ref{equ:transformed_differential_equation}) and substituting the solutions into the boundary conditions in (\ref{equ:boundary}), we have $g(z) = 0$, which contains $1/\lambda_k$ as its roots \cite{gelfand1959calculation}. By deleting useless solutions in $g(z)=0$, we can obtain $f(z) = f(0)\prod_{k=1}^{\infty} (1-z\lambda_k)$ as a direct consequence of the Hadamard's factorization theorem \cite{ahlfors1953complex}.
Then, from (\ref{equ:I_infinite_sum}), we can directly obtain the mutual information without solving every $\lambda_k$. The mutual information can be expressed by 
\begin{equation}
	\begin{aligned}
		I =\log\prod_{k=1}^{+\infty}\left(1+\frac{\lambda_k}{n_0/2}\right)  = \log \frac{f(-\frac{1}{n_0/2})}{f(0)}.
		\label{equ:I_from_f}
	\end{aligned}
\end{equation}

For simple autocorrelation functions like the exponential autocorrelation function we have discussed in the paper, such analysis is much easier than deriving all $\lambda$ to obtain the mutual information. Since any correlation function can be approximated by a function with rational spectrum, this analyzing procedure is extendable to a wide range of scenarios.

Next, we will introduce a special case to show how this analyzing procedure works. The special case is when the electric field has exponential autocorrelation function as $R_E(r,r')=Pe^{-\alpha |r-r'|}$.  We choose this special case not only because it is a very simple case where a stochastic process has a rational spectrum, but also because it has some practical meanings. For example, it is a widely used model when analyzing the scattering field from a surface \cite{franceschetti2006scattering}. If some surfaces in the channel scatter the field, the autocorrelation function may behave like exponential functions.
 The parameter $P\,{\rm [V^2/m^2]}$ determines the power of the received field, and the parameter $\alpha \,{\rm [1/m]}$ controls the correlation of the received field. If we want to solve the eigenvalues of the integral equation (\ref{equ:integral_function}), we can obtain the solutions from \cite{zhujieao} as 
		$\lambda_k = \frac{2\alpha P}{\alpha^2+\omega_k^2}$ and 
		$\phi_k(r) = \frac{1}{Z_k}(\omega_k{\rm cos}(\omega_kr)+\alpha {\rm sin}(\omega_kr))$,
where $Z_k$ are the normalization constants to ensure the orthonormality \eqref{equ:orthogonality}, and the resonant frequencies $\omega_k$ \cite{zhujieao} are the positive solutions to the transcendental equation 
\begin{equation}
\begin{aligned}
2{\rm arctan}(\omega_k/\alpha)= k\pi -\omega_k L,\; k\geq 1,\; k\in\mathbb{Z}.
\end{aligned}
\end{equation}

Although we can obtain $\lambda_k$, we will follow the approach in \cite{gelfand1959calculation} to derive the mutual information and use the $\lambda_k$ derived to verify the correctness. According to (\ref{equ:transformed_differential_equation}) we have 
\begin{equation}
	\phi^{''}(r)-(\alpha^2-\frac{2P\alpha}{\lambda})\phi(r) = 0.
	\label{equ:phi_differential_equation}
\end{equation}

According to (\ref{equ:boundary}) we have the boundary conditions $\phi^{'}(0) - \alpha \phi(0) = 0$ and $\phi^{'}(L) + \alpha \phi(L) = 0$. By solving (\ref{equ:phi_differential_equation}) and substituting it into the boundary conditions, we have 
	$g(z)=2\alpha \omega \cos(\omega L)-(\omega^2-\alpha^2)\sin(\omega L) = 0$, 
where 
	$\omega^2=\frac{2\alpha P}{\lambda} -\alpha^2 = 2\alpha P z-\alpha^2$.

Since $z = \alpha/(2P)$ is a root of $g(z)$ but an exceptional root of the boundary conditions, we remove it from $g(z)$ to construct another entire function $f(z)$ directly from $g(z)$. The $f$ function has the following form
	$f(z) = g(\sqrt{2\alpha P z - \alpha^2})/\sqrt{2\alpha P z - \alpha^2}$.
 It can be proved that all the zeros of $g(\omega)$ lie on the real axis, so all zeros of the entire function $f(z)$ are given by $1/\lambda_k$, arranged in the ascending order on the positive real axis. 

Therefore, from (\ref{equ:I_from_f}), the mutual information is then explicitly expressed as 
\begin{equation}
	\begin{aligned}
	 I &= \log \Bigg(\cosh(\alpha L \sqrt{1+4P/n_0\alpha}) + \frac{1+2P/n_0\alpha}{\sqrt{1+4P/n_0\alpha}}\\&~~~~\sinh(\alpha L \sqrt{1+4P/n_0\alpha})\Bigg) - \alpha L.
	 \label{equ:explicit_MI_expression}
	\end{aligned}
\end{equation}

This analyzing scheme for the linear transceivers can be extended to other scenarios, e.g., circular transceivers. The integral problem can be transferred from (\ref{equ:integral_function}) to
\begin{equation}
	\begin{aligned}
	\lambda_k \phi_k(\theta') &= \int_{0}^{\theta_0}R_E(\theta,\theta')\phi_k(\theta){\rm d}\theta; k>0,k \in {\mathbb N}, \\&~~~~\theta \in [0,\theta_0],\theta^{'} \in [0,\theta_0],
	\end{aligned}
	\end{equation}
	where $\theta_0 \in (0,2\pi)$.
	The assumption that $\theta_0 < 2\pi$ is to avoid the boundary condition $\phi^{(n)}(0) = \phi^{(n)}(2\pi)$. When $\theta_0<2\pi$, the analyzing procedure of this model has no difference from the procedures mentioned above. The mutual information of the scenario when $\theta_0 = 2\pi$ can be viewed as the limitation $\underset{\theta \rightarrow 2\pi}{\rm lim}I_\theta$.

\subsection{General expression based on the Fredholm determinant}

In the previous section, fortunately, we have found an explicit expression of the mutual information between a pair of correlated random fields. The restriction on the random fields is that they should be the observation of stationary random fields in a spatial region, which ensures the existence of the spectrum of the random fields.   
However, in general, the condition that $R_E(r,r')$ only relies on $r-r'$ does not exist.
Therefore, we cannot always find a simple closed-form analytic function whose zeros are exactly all the eigenvalues of a general autocorrelation operator. 
But such an analytic function always exists mathematically, which can be explicitly constructed from the {\it Fredholm determinant} \cite{simon2005trace} of a bounded invertible operator in the form ${\bf 1}+T$.
In this section, we will employ the Fredholm determinant to provide an analytic solution to the mutual information between the fields at the source and the destination, which equals the result from Mercer expansion but has good analytical properties and reliable numerical calculation schemes. 

Let $G$ be the group of bounded operators of the form ${\bf 1}+T$ on a Hilbert space $H$, where $T$ is a trace-class operator \cite{brislawn1988kernels}, and $\bf 1$ is the identity operator. The Fredholm determinant of ${\bf 1}+T\in G$ is defined by the following infinite series:
\begin{equation}
	\det({\bf 1}+T):=\sum_{k=0}^{+\infty} {\rm Tr}(\Lambda^k T),
\end{equation}
where $\Lambda^k T$ is the $k$-th exterior power \cite{simon1977notes} of the bounded operator $T$ on $\Lambda^k H$, and $\Lambda^k H$ denotes the $k$-th exterior product of $H$. Thus, an analytic function $f(z):=\det({\bf 1}+zT)\, ,z\in \mathbb{C}$ can be naturally constructed, from which the mutual information can be evaluated. 
\begin{remark}
	\label{remark_3}
	The Fredholm determinant $\det(\cdot)$ is a homomorphism of $G$ into the multiplicative group of the complex numbers $\mathbb{C}$, i.e., $\det(AB)=\det(A)\det(B), \;\forall A,B\in G$, which is similar to the determinant of a square matrix. Furthermore, suppose the trace-class operator $T$ has eigenvalues $\lambda_k$, then $\det({\bf 1}+T)=\prod_k (1+\lambda_k)$. 
\end{remark}

From the above remark, we can conclude that the Fredholm determinant is a nearly-perfect parallel of the matrix determinant. 
Similar to the fact that the ergodic MIMO capacity can be expressed in a determinant form, the mutual information between continuous regions can also be expressed by the Fredholm determinant.
Specifically, starting from \eqref{mercer_capacity} which is obtained by using Mercer expansion, the mutual information with finite-length destination has already been expressed in the infinite product form
		$I={\rm log}\prod_{k=1}^{+\infty}\left(1+\frac{\lambda_k}{n_0/2}\right)$.
Now we try to express $I$ by the Fredholm determinant of some operator. Since the autocorrelation function $R_E(r,r')$ can be considered as an integral operator $T_E$:
\begin{equation}
	\begin{aligned}
		(T_E\phi)(r) = \int_{0}^{a}R_E(r,r')\phi(r'){\rm d}r',
	\end{aligned}
\end{equation}
according to {\bf Remark \ref{remark_3}}, we can take the Fredholm determinant of the operator $({\bf 1}+T_E/(n_0/2))$ to compute the infinite product $\prod_k (1+\lambda_k/(n_0/2))$, where $\{\lambda_k\}$ are the eigenvalues of $T_E$:
\begin{equation}
	\begin{aligned}
		I=\prod_k\left(1+\frac{\lambda_k}{n_0/2}\right)={\rm log}\,{\det}\left({\bf 1}+\frac{T_E}{n_0/2}\right).
	\end{aligned}
	\label{Fdeterminant}
\end{equation}
Note that the eigenvalues $\lambda_k$ of $T_E$ are equivalent to the eigenvalues of the eigen problem \eqref{equ:integral_function}, and the mutual information $I$ is exactly the function value of the induced analytic function $f(z):=\det({\bf 1}+zT_E)$ evaluated at $z=1/(n_0/2)$. 

Our analysis in the following parts is based on Fredholm determinant for operator $T$ with kernels $K(r,r')$, which correspondes to the linear transceivers. However, the expression of Fredholm determinant and the related analyzing schemes are not restricted to this scenario. The linear transceivers can be naturally extended to rectangular transceivers or other shapes, since the Fredholm determinant is well defined for any trace class operator $T$~\cite{simon2005trace}.

\subsection{Numerical calculation method of Fredholm determinant}
Since the Fredholm determinant is a widely-used tool in physics \cite{wheeler1937mathematical}, its numerical properties have been thoroughly studied. As a result, we can provide a numerical calculation method for the mutual information $I$ and the corresponding convergence analysis according to \cite{bornemann2010numerical}. Inspired by the numerical integral method, the operator $T_E$ can be approximated by discrete summation 
\begin{equation}
\begin{aligned}
(T_E\phi)(r) := \int_{0}^{a}K(r,r')\phi(r'){\rm d}r' \approx \sum_{j=1}^{m} w_j K(r,r'_j) \phi(r'_j),
\end{aligned}
\end{equation}
where $K=R_E: [0,L]^2\to \mathbb{C}$ is called the kernel of the integral operator $T_E$, and the integral equation can then be discretized into
\begin{equation}
\begin{aligned}
\sum_{j=1}^{m} w_j K(r_i,r_j)\phi_k(r_j) = \lambda_k \phi_k(r_i).
\end{aligned}
\end{equation}
When $K(r_i,r_j)=\delta(r_i,r_j)$, the identity operator ${\bf 1}$ is discretized into an identity matrix
${\bf 1} \sim  \sum_{j=1}^{m} {\mathbbm 1}_{i=j}$.
The Fredholm determinant can then be approximated by
\begin{equation}
\begin{aligned}
I&\approx {\rm log}{\rm det}\left({\mathbbm 1}_{i=j}+\frac{w_j K(r_i,r_j)}{n_0/2}\right)_{i,j=1}^{m}
\\&={\rm log}{\rm det}\left({\mathbbm 1}_{i=j}+\frac{w_{j}^{1/2} K(r_i,r_j)w_{j}^{1/2}}{n_0/2}\right)_{i,j=1}^{m}.
\end{aligned}
\label{matrixdeterminant}
\end{equation}

The convergence of this discretization method is proven in \cite{bornemann2010numerical}, which is dependent on the way of discretizing the integral operator $T_E$. For example, Gauss-Legendre quadrature has been proved to have satisfactory convergence properties, and achieves the highest degrees of precision. A Gauss-Legendre quadrature with $n$ points has degrees of precision $2n-1$ and order $2n$. The Gauss-Legendre quadrature on the integration interval $[-1,1]$ is based on the  Legendre orthogonal polynomials 
	$P_0(x) :=  1$ and
	$P_n(x) := \frac{1}{2^n n!}\frac{{\rm d}^n}{{\rm d}x^n}\left[ (x^2-1)^n \right]$. 

The Gauss points $x_1,\dots,x_n$ are the zeros of $P_{n}(x)$. The weights $A_1,\cdots,A_n$ can be derived as follows
\begin{equation}
\begin{aligned}
A_k = \frac{2}{n} \frac{1}{P_{n-1}(x_k)P^{'}_{n}(x_k)}.
\end{aligned}
\end{equation}
The convergence of (\ref{matrixdeterminant}) can be derived as
\begin{equation}
\begin{aligned}
\left| Q(f) - \int_0^{a}f(x){\rm d}x  \right| \leqslant c_k a^{k+1} v^{-k} \left\|f^{(k)} \right\|_{L^{\infty}(0,a)},
\end{aligned}
\end{equation}
for $f \in C^{k-1,1}([0,a])$, quadrature rule $Q$ of order $v \geqslant k$ with positive weights. The reference \cite[{\bf Theorem 6.2}]{bornemann2010numerical} shows that
\begin{equation}
\begin{aligned}
\left| d_Q(z) - d(z)  \right| \leqslant c_k 2^k a^{k} v^{-k} \Phi(\left|z\right| a \left\|K \right\|_k),
\end{aligned}
\label{capacity_converge}
\end{equation}
where $d(z) := \det(I+zT)$, $K \in C^{k-1,1}([0,a]^2)$, quadrature rule $Q$ of order $v \geqslant k$ with positive weights,
$\left\|K \right\|_k = \underset{i+j \leqslant k}{\rm max}\left\| \partial_1^i \partial_2^j K \right\|_{L^{\infty}}$ and 
$\Phi(z) = \sum_{n=1}^{+\infty}\frac{n^{(n+2)/2}}{n!}z^n$.

The convergence of the numerical calculation scheme guarantees the reliability of the scheme. Therefore, the Fredholm determinant is not only a mathematical representation of the mutual information, but also provides a useful tool to calculate the mutual information that can be obtained from the received noisy field. 

\section{Mutual Information with Finite-Length Transceivers under Colored Noise Field}
\label{sec:colored noise field}
In the above section, we have discussed the mutual information between finite-length source and destination under the assumption of a white noise field, which is a common model adopted by recent works. However, the white noise model is only a simplified model that comes from the assumption of traditional MIMO modeling. In this section, we consider other mechanisms which have been analyzed in Section \ref{sec:spatial analysis} that may contribute to the noise, and analyze the mutual information of the proposed model.  A general numerical calculation procedure is proposed of the mutual information, which can be viewed as the extension of the numerical scheme in Section \ref{sec: upperbound}.

\subsection{Mutual information based on operator analysis}
In Section.~\ref{random_field_modeling_of_noise}, we have derived a non-white noise model, which means that the noise field has non-zero spatial correlation coefficients. Following the same procedure, we assume that the covariance kernel of the noise field is $K_N(r,r')\neq P \delta(r-r')$, and the corresponding operator is $T_N:\mathscr{L}^2(V_R)\to\mathscr{L}^2(V_R)$. The operator of the received noisy field $Y=E+N$ is denoted by $T_Y$.  Here we can perform Mercer expansion on the electric field $E(r)$ and the noise field $N(r)$ similar to Section \ref{sec: upperbound}, which leads to $R_E(r,r') = \sum_k \lambda_k\phi_k(r)\phi_k^{*}(r')$, $E(r) = \sum_k\xi_{E,k} \phi_k(r)$, $R_N(r,r')=\sum_k \lambda^{'}_k \psi_k(r)\psi_k^{*}(r')$ and $N(r) = \sum_k\xi_{N,k} \psi_k(r)$. The received field $Y(r)$ can also be decomposed as $Y(r) = \sum_k \xi_{Y,k} \varphi_k(r)$. However, different from the white noise case where $\psi(r)$ can be arbitrary orthogonal bases, the expansion of non-white noise is confirmed, which means that the decomposition of $E(r)$, $N(r)$ and $Y(r)$ will lead to different sets of eigenfunctions and the mutual information can not be directly obtained like (\ref{mercer_capacity}). 

Inspired by \cite{gelfand1959calculation}, we can construct two spaces $H_1$ and $H_2$ which are extended by $\xi_{E,k}$ and $\xi_{Y,k}$ separately. The inner product in these spaces is defined as the cross-correlation between two random variables in the space. The mutual information between $E(r)$ and $Y(r)$ is actually the difference between the space $H_1$ and $H_2$ and is dertermined by the angle between them according to \cite{gelfand1959calculation}. Specifically, the mutual information can be evaluated by finding new sets of bases $\hat{\xi}_{E,k}$ and $\hat{\xi}_{Y,k}$ in $H_1$ and $H_2$ which satisfy $\mathbb{E}[\hat{\xi}_{E,k_1}\hat{\xi}_{Y,k_2}^{*}] = 0$, $\mathbb{E}[\hat{\xi}_{E,k_1}\hat{\xi}_{E,k_2}^{*}] = 0$ and $\mathbb{E}[\hat{\xi}_{Y,k_1}\hat{\xi}_{Y,k_2}^{*}] = 0$ when $k_1\neq k_2$. These $\hat{\xi}$ can be viewed as the linear combinations from $\xi$ decomposed from $E(r)$ and $Y(r)$. Moreover, they correspond to the projection operators $B_1 = P_1P_2$ and $B_2 = P_2P_1$, where $P_1$ is the projection onto the space $H_1$ and $P_2$ is the projection onto the space $H_2$. We have 
\begin{equation}
	\begin{aligned}
I &= -\sum_k {\rm log} (1- \frac{\mathbb{E}[\hat{\xi}_{E,k}\hat{\xi}_{Y,k}^{*}]^2}{\mathbb{E}[\hat{\xi}_{E,k}\hat{\xi}_{E,k}^{*}]\mathbb{E}[\hat{\xi}_{Y,k}\hat{\xi}_{Y,k}^{*}]}  ) \\&= -{\rm log}{\rm det}({\bf 1} - P_1P_2) = -{\rm log}{\rm det}({\bf 1} - P_2P_1).
	\end{aligned}
\end{equation}

Since $P_1$ is the projection operator, we have $\mathbb{E}[P_1\xi_{E}*\xi_{E}^*] = \mathbb{E}[\xi_{Y}\xi_{E}^*]$. By defining the operator $T_D$ and $T_{D^{'}}$ which represent the mutual correlation between $E(r)$ and $Y(r)$, we can obtain $P_1$ and $P_2$, thus obtaining the mutual information. Here we define $T_D$ by $(T_D \psi)(r):=\int_{V_R}\mathbb{E}[Y(r)E^*(r')]\psi(r'){\rm d}r'$ and $T_{D^{'}}$ by $(T_{D^{'}} \psi)(r):=\int_{V_R}\mathbb{E}[Y^*(r)E(r')]\psi(r'){\rm d}r'$. They can also be represented by infinite-size matrix to be easier to understand, as $T_D \sim  \left\| {\mathbb E} \left[  \langle \phi_j |{Y} \rangle \langle E|\phi_i\rangle\right] \right\|_{i,j=1}^{m}$. Through this approach we have $P_1T_E = T_D$ and $P_2T_Y = T_{D^{'}}$, which means $P_1 = T_D T_E^{-1}$ and $P_2 = T_{D^{'}}T_Y^{-1}$. Therefore we have
\begin{equation} 
	\begin{aligned}
		I(J;Y) &= I(E;Y) = -{\rm log}{\rm det}({\bf 1}-T_DT_E^{-1}T_{D'}T_Y^{-1}) \\&= -{\rm log}{\rm det}({\bf 1}-T_{D'}T_Y^{-1}T_DT_E^{-1}),
	\end{aligned}
	\label{mutual_information_operator}
\end{equation}  

This result inspired by \cite{gelfand1959calculation} is about stochastic process over the real line and the sampling of it is a single variable, which coincides with the model of parallel linear transceivers that we used. It is necessary to mention that such expression based on operator is not restricted to this simplified scenario, but can be extended to random field over arbitrary manifolds and the sampling of it can be a random vector, which has theoretical bases from \cite{gangolli1963wide}. By using random field on two-dimensional surfaces we can extend the model from linear transceivers to rectangular or circular transceivers. By using random vector as the sampling result we can extend the scalar wave field to three-dimensional wave field.

In our assumption, the noise field $N$ is independent of the noiseless electric field $E$, 
which means $T_D=T_{D'}= T_E$. The mutual information in (\ref{mutual_information_operator}) is then simplified to 
\begin{equation}
	\begin{aligned}
		I(E;Y) &= -{\rm log}{\rm det}({\bf 1}-T_DT_E^{-1}T_{D'}T_Y^{-1})\\&=-{\rm log}{\rm det}({\bf 1}-T_E(T_E+T_N)^{-1}).
		\label{mutual_information_operator_1}
	\end{aligned}
\end{equation}
In the special case when the noise field is assumed to be a white Gaussian random field, the kernel corresponding to the noise operator is $K_N(r,r') = \frac{n_0}{2}\delta(r-r')$. Since for arbitrary $\phi(r)$ the following integral equation holds
		$\int_0^{a}K_N(r,r')\phi(r'){\rm d}r' = \frac{n_0}{2}\phi(r)$,
we have
		$T_N \ket{\phi} = \frac{n_0}{2} \ket{\phi}$.
Therefore, the difference between $T_N$ and the identity operator $\bf 1$ is only a constant factor, i.e., we can represent $T_N$ by $T_N=\frac{n_0}{2}{\bf 1}$. The mutual information in (\ref{mutual_information_operator_1}) can be further simplified to 
\begin{equation}
	\begin{aligned}
		I(E;Y) &=-{\rm log}{\rm det}({\bf 1}-T_E(T_E+T_N)^{-1})
		 \\&= {\rm log}{\rm det}\left({\bf 1}+\frac{T_E}{n_0/2}\right),
		\label{mutual_information_operator_2}
	\end{aligned}
\end{equation}
which coincides with the formula derived in (\ref{Fdeterminant}). Therefore, the mutual information under white noise model can be viewed as a special case of the result in this section.

\subsection{Numerical calculation of the mutual information}
As shown in the previous subsection, we can express the mutual information between the electric field and the received noisy field by (\ref{mutual_information_operator_1}). Now we want to find a numerical algorithm to calculate the mutual information, and show how the noise field affects the information we can obtain from the received field.

In Section \ref{sec: upperbound}, we have clarified that the numerical approximation of the mutual information converges under the white-noise scenario. 
For the colored noise scenario, if we view the operator $T_E(T_E+T_N)^{-1}$ as a whole, we can directly utilize the conclusion of convergence in the case of white noise. However, we can not directly discretize $T_E(T_E+T_N)^{-1}$ because the inverse operator $(T_E+T_N)^{-1}$ is hard to obtain. Therefore, we discretize $T_E$ and $T_E+T_N$ separately instead of directly discretizing $T_E(T_E+T_N)^{-1}$. The algorithm of the discretization is as {\bf Algorithm \ref{alg:numerical}}, where the kernel of $T_N$ equals $R_{N,{\rm non-white}}(r,r')+\sigma_{\rm white}^2\delta(r-r')$.

\begin{remark}
	The discretization algorithm of the mutual information not only has the mathematical meaning but also has physical correspondence. Since $-{\rm logdet}({\bf I}-{\bf K}_{\rm sep})$ in Algorithm \ref{alg:numerical} equals ${\rm log}\,{{\rm det}({\bf K}_{E}+{\bf K}_{N})/{\rm det}({\bf K}_{N})}$, which is the form of the mutual information of multiple Gaussian channel model. If we express ${\bf K}_{E,i,j}$ and ${\bf K}_{N,i,j}$ by $\mathbb{E}\left[ \int_{a_{i-1}}^{a_i}E(x){\rm d}x \int_{a_{j-1}}^{a_j}E^{*}(y){\rm d}y  \right]$ and $\mathbb{E}\left[ \int_{a_{i-1}}^{a_i}N(x){\rm d}x \int_{a_{j-1}}^{a_j}N^{*}(y){\rm d}y  \right]$ respectively, it is easy to find that the mutual information after discretization can be viewed as using several patch antennas in the destination region which receives the integral of the electric field. Therefore, the convergence of the numerical approximation of the mutual information shows how the discrete destinations approach the continuous one. Here we use Gauss-Legendre quadrature rule because it converges faster than the equally-spaced trapezoid or rectangular quadrature rules. 
\end{remark}

\begin{algorithm}[t]
	\caption{Numerical calculation scheme for the mutual information} 
	\label{alg:numerical}
	\hspace*{0.02in} {\bf Input:} 
	\\
	\hspace*{0.4in} $R_E(r,r')$   \Comment{the kernel of the operator $T_E$} \\
	\hspace*{0.4in} $R_{N,{\rm non-white}}(r,r')$    \\
	\hspace*{0.4in} $\sigma_{\rm white}^2$   \\
	\hspace*{0.4in} $a$   \Comment{the length of the destination} \\
	\hspace*{0.4in} $m$  \Comment{the size of the discretized matrix} \\
	\hspace*{0.02in} {\bf Output:} \\
	\hspace*{0.4in} $ I_{\rm approx} $  \Comment{approximation of the mutual information} 
	\begin{algorithmic}[1]
		\State $(w_1^{m},r_1^{m})$ $\leftarrow$ Quadrature$(0,\,a,\,m)$ 
		\State $a_0$ $\leftarrow$ 0
		\For {$1 \leq i \leq m$}
		\State $a_i$ $\leftarrow$ $a_{i-1}+w_i$
		\EndFor
		\For {$1 \leq i \leq m$}
		\For {$1 \leq j \leq m$}
		\State ${\bf K}_{E,i,j}$ $\leftarrow$ $\int_{a_{i-1}}^{a_i}\int_{a_{j-1}}^{a_j}K_E(x,y){\rm d}y{\rm d}x \approx  w_i w_j R_E(r_i,r_j)$  
		\State ${\bf K}_{N,i,j}$ $\leftarrow$ $\int_{a_{i-1}}^{a_i}\int_{a_{j-1}}^{a_j}K_N(x,y){\rm d}y{\rm d}x\approx w_i w_j R_{N,{\rm non-white}}(r_i,r_j) +w_i\sigma^2_{\rm white}{\mathbbm 1}_{i=j}$
		\EndFor
		\EndFor
		\State ${\bf K}_{\rm sep}$ $\leftarrow$ ${\bf K}_{E}({\bf K}_{E}+{\bf K}_{N})^{-1}$
		\State $I_{\rm approx}$ $\leftarrow$ $-{\rm log}{\rm det}({\bf I}-{\bf K}_{\rm sep})$
		
		\State \Return $ I_{\rm approx} $
	\end{algorithmic}
\end{algorithm}

The algorithm above that separately discretize $T_E$ and $T_E+T_N$ is used to numerically calculate the information obtained from the received electric field under non-white noise. The accuracy is guaranteed by the quadrature rule. Based on the numerical scheme, we will provide some numerical results of the mutual information. Here we assume that the source and destination both have length $L$, the current density $J(s)$ on the source has the autocorrelation function $R_J(r,r')=\delta(r-r')$, which means that no channel state information at the transmitter (CSIT) is available. Then, the electric field on the destination can be derived by
\begin{align}
	R_E(r,r') &= \int_{0}^{L} \int_{0}^{L} g(r,s)\delta(s-s') g^{*}(r',s'){\rm d}s{\rm d}s' \notag
	\\&= \int_{0}^{L} g(r,s) g^{*}(r',s){\rm d}s.
\end{align}
 The noise field is considered a mixture of the measurement noise and the radiation interference. We denote the mixed noise field\footnote{Here the unit of $\sigma_1^2$ is $[{\rm V}^2/{\rm m}]$ and the unit of $\sigma_2^2$ is $[{\rm V}^2/{\rm m}^2]$. These parameters are used to keep the unit of $N(r)$ to be $[{\rm V}/{\rm m}]$. The units of $\sigma_1^2$ and $\sigma_2^2$ are different because the measurement noise comes from unideal measurement or data processing steps, while the radiated interference is strongly correlated to the wavelength of the electromagnetic field.} by $N(r) = \sigma_1 N_{\rm mea}(r)+\sigma_2 N_{\rm int}(r)$ and its autocorrelation function by $R_N(r,r')$. According to \eqref{eqn:parallel_direction}, we have
 \begin{equation}
	\begin{aligned}
	&~~~~R_N(r,r') \\&= \sigma_1^2 \delta(r-r') +2 \sigma_2^2 \left(\frac{{\rm sin}(\kappa r-\kappa r')}{(\kappa r-\kappa r')^3}-\frac{{\rm cos}(\kappa r-\kappa r')}{(\kappa r-\kappa r')^2}\right).
	\end{aligned}
\end{equation}

\begin{figure}
	\centering 
	\includegraphics[height=6cm, width=8cm]{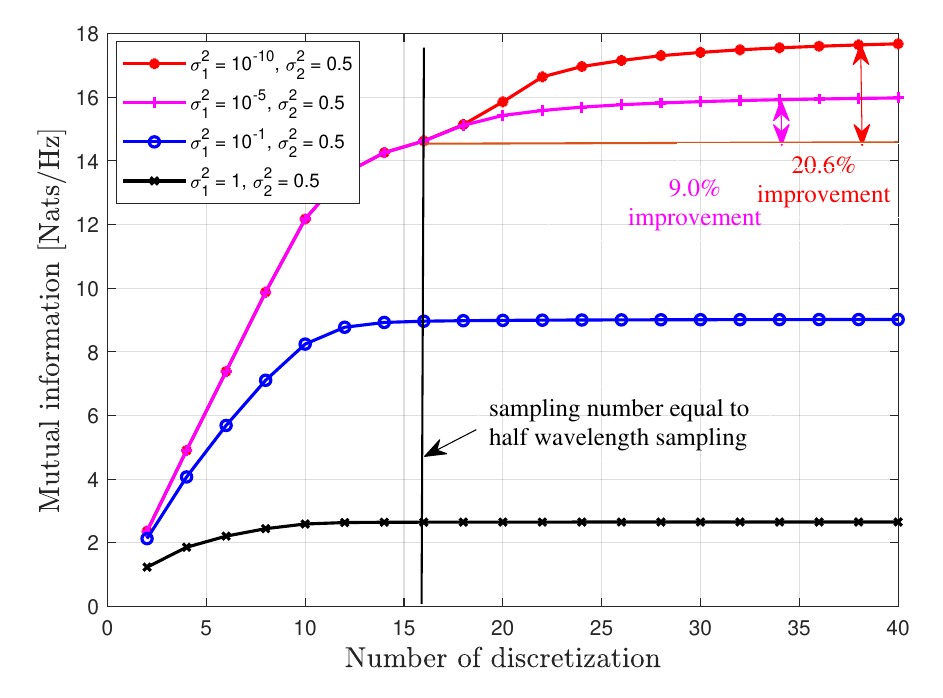} 
	\caption{Mutual information $I(L)$ in \eqref{mutual_information_operator_1} calculated by applying {\bf Algorithm \ref{alg:numerical}}. The $x$-axis represents the number of discretization points, and the $y$-axis is the mutual information measured in nats. The mutual information value increases as the number of discretization points increases, and finally converges to the continuous-space mutual information defined by the supremum in \eqref{equ:mutual_info_definition}. } 
	\label{fig:convergence}
\end{figure}

Now we set the length $L$ of the source and destination to $2\,{\rm m}$ and the wavelength $\lambda$ of the electromagnetic wave to $0.25\,{\rm m}$. The distance between the source and the destination is fixed to $1\,{\rm m}$. Therefore, the half-wavelength sampling on the destination has $16$ sampling points. The mutual information obtained from the discretization is expressed in Fig. \ref{fig:convergence}. In this figure, the power of the radiation interference is fixed to $\sigma_2^2 = 0.5\;[{\rm V}^2/{\rm m}^2]$, and the measurement noise is flexible. It is shown that when the power of the measurement noise is large, the mutual information converges quickly with the increasing number of discretization points. 
For example, when $\sigma_1^2 = 1\;[{\rm V}^2/{\rm m}]$ or $\sigma_1^2=10^{-1}\;[{\rm V}^2/{\rm m}]$, the information obtained from the continuous destination is almost the same as the information obtained from the 16-points discretization. 
However, when $\sigma_1^2\;[{\rm V}^2/{\rm m}]$ is very small, which means that the radiation interference plays a dominant role in the noise model, the gap between the continuous case and the discretized case becomes obvious. When $\sigma_1^2 = 10^{-5}\;[{\rm V}^2/{\rm m}]$, the information obtained from the continuous destination is $9\%$ larger than the information obtained from the 16-points discretized destination. When $\sigma_1^2 = 10^{-10}\;[{\rm V}^2/{\rm m}]$, the information obtained from the continuous destination is $20.6\%$ more than that obtained from the 16-points discretized destination. 

Fig.~\ref{fig:convergence} also demonstrates that the half-wavelength sampling is strictly suboptimal, since $16$ discretization points fail to harvest all the mutual information provided by \eqref{mutual_information_operator_1}. Furthermore, the extent of such suboptimality depends on the ratio $\sigma_1^2/\sigma_2^2$ of the white noise power with respect to the non-white noise power. This is because the colored interference noise is more ``structural'', and lives in a lower-dimensional noisy subspace, while the white noise projects equally into any subspace. Thus, a smaller portion of white noise leads to more clean signal subspaces, which further results in more mutual information.


\section{Mutual Information with Infinite-Length Transceivers}
\label{sec:two infinite}
In the above sections, we have analyzed the mutual information between finite-length transceivers and provided its expression based on the Fredholm determinant. To further strengthen the connection between the model of parallel linear transceivers and classical model in information theory, we introduce an ideal model in which the transceivers are of infinite length. A strong correspondence between this model in the spatial dimension and the classical signal model in the time dimension is built. Based on this correspondence, we utilize the analyzing methods in classical information theory to provide the mutual information and the capacity for this scenario.

\subsection{Connection between the infinite-length transceivers and classical models in information theory}
Since we extend the finite-length transceivers to infinite-length transceivers in our new ideal model, we have 
	$E(r) = \int_{-\infty}^{+\infty} {g} (r,s)J(s){\rm d}{s}$. 
The main difference between this model and the model with finite-length transceivers is that the current density and the electric field become random processes on the infinite-length spatial dimension, which is similar to the channel model $Y(t)=\int_{-\infty}^{+\infty} {h} (t,\tau)X(\tau){\rm d}{\tau}$ in classical information theory. 
Similar to the classical information theory which often assumes that $X(\tau)$ and $Y(t)$ are stationary random processes, we assume that, the current density $J$ and electric field $E$ are also stationary stochastic processes in the space domain. The spatial stationary means that, $\mathbb E[J(s)J^{*}({s}')]$ only depends on $\Delta s={s}'-s$, and we can introduce the autocorrelation function as
$R_{J}(\Delta s) = {\mathbb E}[J(s)J^{*}({s}')]$
and
	$R_{E}(\Delta r)={\mathbb E}[E(r)E^{*}(r')]=\int_{-\infty }^{+\infty }\int_{-\infty }^{+\infty }g(r,s)R_{J}(\Delta s)g^{*}(r',s'){\rm d}s{\rm d}s'$.

The SSD can then be derived as 
\begin{equation}
S_{J}(\kappa) = \frac{1}{\sqrt{2\pi}} \int_{-\infty }^{+\infty }R_J(\Delta s)e^{-{\rm j} \kappa \Delta s} {\rm d} \Delta s.
\end{equation}

The SSD of the electric field can then be derived by the SSD of the current density via the Fourier transform of the Green's function $G(\kappa)$:
\begin{equation}
\begin{aligned}
S_{E}(\kappa)=2\pi S_{J}(\kappa)|G(\kappa)|^2.
\end{aligned}
\label{SE_with_SJ}
\end{equation}

\subsection{Spatial spectrum analysis of the communication between transceivers}
In this subsection, we will reveal how some important parameters affect the communication quality of the system with continuous source and destination from $G(\kappa)$, which is the Fourier transform of the Green function. We introduce \textbf{Lemma 1} to derive a closed-form solution for $G(\kappa)$.

\begin{lemma} \label{thm:Green}
	The Fourier transform of the Green function in (\ref{equ:Green_g}) can be expressed by $G(\kappa)=F_1(\kappa)*F_2(\kappa)$, where
	\begin{equation}
	\begin{aligned} 
	&{F}_1(\kappa) = \frac{-{\rm j}{ {Z_0} {d^2}}}{{4\pi  \lambda }}\left\{ 
	\begin{array} {lcr}
	\frac{\pi}{2} {\rm j} {J_0}(dm ) - \frac{\pi}{2} {Y_0}(dm ), & [0 < \kappa < \frac{{2\pi }}{\lambda }]\\ 
	{K_0}{\rm{(}}dm {\rm{) }}, & [\kappa > \frac{{2\pi }}{\lambda }] 
	\end{array} 
	\right.
	\end{aligned}
	\end{equation}
	with $m=\sqrt {|{{(\frac{{2\pi }}{\lambda })}^2} - {\kappa^2}|} $, and
	\begin{equation}
	\begin{aligned}
	F_2(\kappa) =~ & d{\sqrt {\frac{\pi }{2}} {e^{ - d\left| \kappa \right|}}}+\frac{{\rm j}d\lambda}{2\pi}\sqrt {\frac{2}{\pi }} \left| \kappa \right|{K_1}(d\left| \kappa \right|) \\&-\frac{{\rm j}\lambda}{\pi}\Big[\sqrt {\frac{2}{\pi }} {K_0}(d\left| \kappa \right|) - d\sqrt {\frac{2}{\pi }} \left| \kappa \right|{K_1}(d\left| \kappa \right|)\Big] \\&-(\frac{\lambda}{2\pi})^2\frac{1}{2d}\sqrt {\frac{\pi }{2}} (1 + \left| {d\kappa} \right|){e^{ - \left| {d\kappa} \right|}}
	\\&+(\frac{\lambda}{2\pi})^2\Big[\sqrt {\frac{\pi }{2}}\frac{2{ {e^{ - d\left| \kappa \right|}}}}{d} - \frac{1}{d}\sqrt {\frac{\pi }{2}} (1 + \left| {d\kappa} \right|){e^{ - \left| {d\kappa} \right|}}\Big].
	\end{aligned}
	\label{F2kappa_0}
	\end{equation} 
	
\end{lemma}
\begin{IEEEproof}
	See Appendix A.
\end{IEEEproof}

From the Fourier transform of the Green function in \textbf{Lemma 1}, we can find that the distance between the source and destination $d$ deeply affects the behavior of $|G(\kappa)|$. Since $|G(\kappa)|$ depicts how the spectrum of the current on the source affects the spectrum of the electric field on the destination, larger values of $|G(\kappa)|$ will cause higher channel gain in the corresponding wavenumber bands. 

\begin{figure}
	\centering 
	\includegraphics[height=6cm, width=8cm]{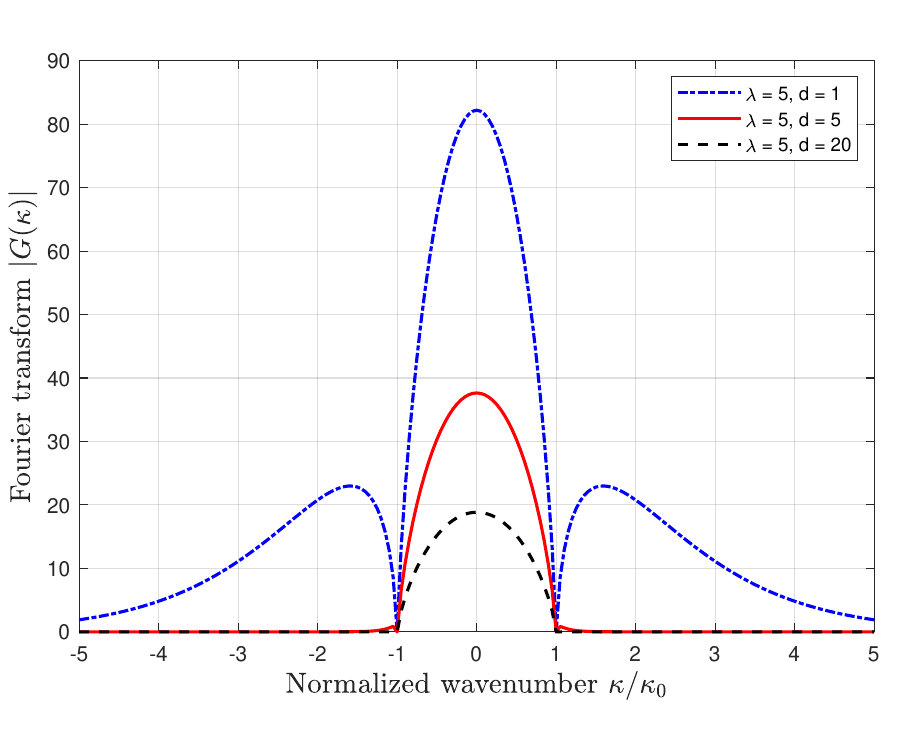} 
	\caption{Fourier transform of the Green function with different distances $d$.} 
	\label{fig:Green_d}
\end{figure}

In Fig. \ref{fig:Green_d}, we plot $|G(\kappa)|$ with different distances $d$ while $\lambda$ is fixed to 5$\,{\rm m}$. 
We find that $|G(\kappa)|$ has a main lobe in the band $[-\kappa_0,\kappa_0]$, which means that the electromagnetic wave is a plane wave. When $\kappa$ equals $0$, the wavefront of the plane wave is parallel to the linear destination. The figure of $G(\kappa)$ has side lobes when $\kappa$ falls out of the band $[-\kappa_0,\kappa_0]$, which means that the corresponding component of the electromagnetic wave is an evanescent wave and vanishes when the distance is larger than the wavelength.
 The channel with obvious side lobes can support broad bandwidth in the wavenumber domain, thus the DoF can be increased. We find that, when $d$ decreases, the main lobe and the side lobes both have more energy. Therefore, the DoF and the gain per degree both increase. From this phenomenon, we conclude that with small $d$ which is comparable with the wavelength, we can get more DoFs besides the power gain per degree, thus improving the channel capacity. This phenomenon is in the reactive near-field region, which only has theoretical meanings and is acceptable to be neglected in the current wireless communication scenarios.

\subsection{Capacity between the transceivers}
We suppose that the noise obeys the spatial additive white Gaussian noise (AWGN) model, which means that the autocorrelation function of the noise $R_N(\Delta s)=\sigma^2\delta(\Delta s)$ and the SSD of the noise is $S_N(\kappa)=\frac{\sigma^2}{\sqrt{2\pi}}$. Here we follow~\cite{jensen2008capacity} to assume that the power constraint\footnote{The radiation power which can be calculated from the Poynting vector is physically more meaningful, but the optimal field distribution is very hard to derive under this power constraint, so here we impose the power constraint on the SSD.} is similar to that of the traditional MIMO model, which constrains the sum of the squares of current excitations on the source antennas. Note that the integral of the square of current in the spatial domain can be transferred to the integral of its spectrum in the wavenumber domain based on the Parseval's theorem. Thus, the power consumption for the source can be bounded by
\begin{equation}
\int_{-\infty}^{+\infty}S_J(\kappa){\rm d}\kappa \leqslant P.
\label{equ:powerconstraint}
\end{equation}

Inspired by the integral form of the Shannon capacity for colored AWGN noise channel \cite{shannon1949communication, Cover1999ElementsInfTheory}, the channel can be split to infinite narrow-band subchannels in the wavenumber domain. For each subchannel, the capacity is
${\rm d}C_{\kappa_1}=\frac{1}{2 \pi}{\rm log}\left(1+\frac{S_E(\kappa_1)}{S_N(\kappa_1)}\right){\rm d}\kappa_1$.
Therefore, the overall capacity is
\begin{equation}
\begin{aligned}
C=\frac{1}{2 \pi}\int_{-\infty}^{+\infty}{\rm log}\left(1+\frac{2\pi S_J(\kappa)}{S_N(\kappa)/|G(\kappa)|^2}\right){\rm d}\kappa.
\end{aligned}
\label{equ:cap_two_infinite}
\end{equation}

To simplify the analysis, we introduce the equivalent noise $N'$ with SSD $S_{N'}(\kappa)=S_N(\kappa)/(2\pi|G(\kappa)|^2)$. To derive the best SSD of the current density $S_J(\kappa)$, we obtain the following \textbf{Theorem 1}.

\begin{theorem} 
	\label{thm:variantion calculus}
	For the overall capacity in (\ref{equ:cap_two_infinite}) and the power constraint for the current density in (\ref{equ:powerconstraint}), the SSD of the best current density obeys $S_J(\kappa) = \left(\frac{1}{2\pi \mu} - S_{N'}(\kappa)\right)^{+}$, where $\mu$ is obtained by the power constraint.
	
\end{theorem}

\begin{IEEEproof} To obtain the optimal SSD $S_J(\kappa)$ that maximizes \eqref{equ:cap_two_infinite}, we apply the variational calculus \cite{gelfand2000calculus}. The Lagrange multiplier $\mu$ can be introduced to take into consideration the power constraint $\int{S_J {\rm d}\kappa}\leqslant P_J$:
	\begin{equation}
	\begin{aligned}
	\mathcal{L}(S_J, \mu)  = & ~\frac{1}{2 \pi}\int_{-\infty}^{+\infty}{\log\left(1+\frac{S_J(\kappa)}{S_{N'}(\kappa)}\right){\rm d}\kappa} \\& - \mu \left(\int_{-\infty}^{+\infty}{S_J(\kappa){\rm d}\kappa}-P_J\right).
	\end{aligned}
	\label{equ:lagrange_function}
	\end{equation}
	Taking the variation of \eqref{equ:lagrange_function} we obtain
	\begin{equation}
	\delta C=\frac{1}{2 \pi}\int_{-\infty}^{+\infty}{\frac{1}{1+\frac{S_J(\kappa)}{S_{N'}(\kappa)}}\frac{\delta S_J(\kappa)}{S_{N'}(\kappa)}{\rm d}\kappa} - \mu \int_{-\infty}^{+\infty}{\delta S_J(\kappa){\rm d}\kappa}.
	\label{equ:variational}
	\end{equation}
	From \eqref{equ:variational} we obtain that the optimal solution for $S_J$ should satisfy
	$\frac{1}{{2 \pi}(S_J(\kappa)+S_{N'}(\kappa))} - \mu \equiv 0$.
	Taking the non-negative condition on $S_J$ into consideration, we obtain
	$S_J(\kappa) = \left(\frac{1}{2\pi \mu} - S_{N'}(\kappa)\right)^{+}$.
\end{IEEEproof}

\begin{figure}
	\centering 
	\includegraphics[height=6cm, width=8cm]{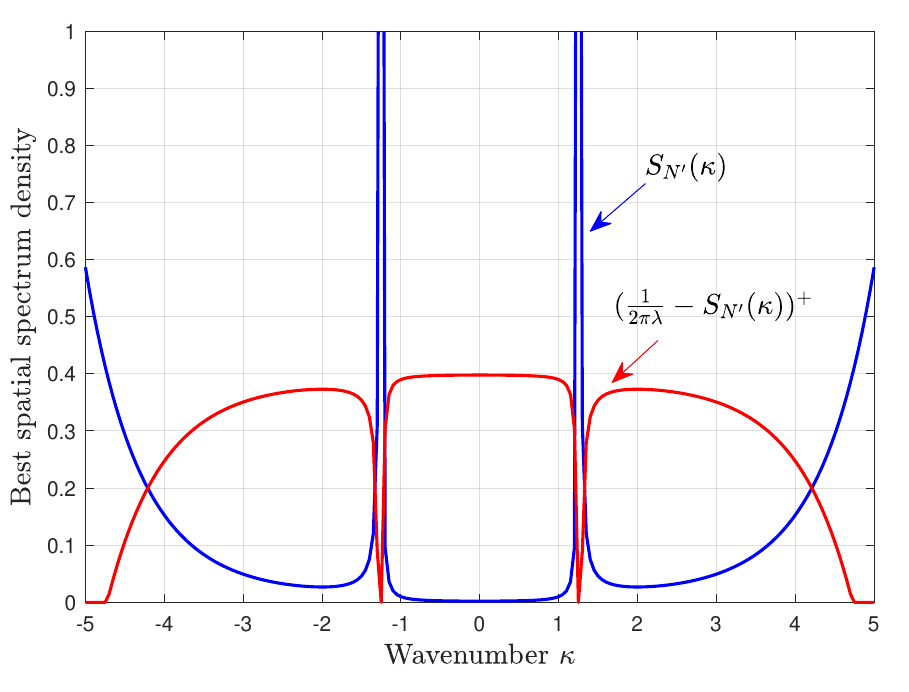} 
	\caption{Optimal SSD of the current density on the source based on variational calculus.} 
	\label{fig:waterfilling}
\end{figure}

For example, we consider a set of parameters where the wavelength $\lambda$ is 5$\,{\rm m}$ and the distance~$d$ between the two parallel lines is 1$\,{\rm m}$.
We use AWGN model with the noise SSD  $S_N(\kappa)=90\,[{\rm V^2/m}]$. The power constraint for the current density $P$ is equal to $3\,[{\rm A^2/m^4}]$. From {\bf Theorem 1} we can derive the best SSD of the source current, as shown in Fig. \ref{fig:waterfilling}. It is worth noticing that the scenario considered here is in the reactive near field, when the wavelength is comparable to the distance between the transceivers. The best spectrum density in the $[-\kappa_0,\kappa_0]$ band is nearly flat. This phenomenon coincides with Fig. 6 in \cite{sanguinetti2022wavenumber}, where discrete Fourier bases are chosen to decompose the channel instead of the continuous Fourier transform we adopted in this section.

\section{Conclusions}
\label{sec:conclusion}	
In this paper, we analyzed the mutual information and the capacity of electromagnetic fields based on random field theory. We first developed the system model of communication between two continuous regions based on random fields. The model based on random fields provided a general framework for solving the problems like mutual information and capacity for EIT. Then, we considered a simplified model with parallel finite-length linear source and destination and analyze the mutual information under white and colored noise fields. Numerical results confirmed the suboptimality of the half-wavelength sampling. Finally, we considered an ideal model with an infinite-length destination and obtained the mutual information and capacity through SSD. Moreover, the gain of DoF and mutual information in the reactive near-field region is shown using SSD.

Further work will focus on the analytical solutions of more general cases like non-linear sources and destinations. The correlation between the current density distribution on the source and the radiated power in the space is also necessary to be explored.



%


\section*{Appendix A \\ Proof of Lemma 1}
For the Green function in (\ref{equ:Green_g}), we split it into the product of two functions $g(x)=f_1(x)f_2(x)$, where
\begin{subequations}
	\begin{align} 
	&{f}_1(x) = \frac{-{\rm j}{ {Z_0}\eta {e^{{\rm j}2\pi \sqrt {{x^2} + {d^2}} /\lambda }}}}{{2\lambda \sqrt {{x^2} + {d^2}} }},\\
	&{f}_2(x) = \frac{{{d^2}}}{{{x^2} + {d^2}}} + \frac{\rm j}{{2\pi \sqrt {{x^2} + {d^2}} /\lambda }}\frac{{{d^2} - 2{x^2}}}{{{x^2} + {d^2}}} \\&~~~~~~~~~~- \frac{1}{{{{(2\pi /\lambda )}^2}({x^2} + {d^2})}}\frac{{{d^2} - 2{x^2}}}{{{x^2} + {d^2}}}\notag.
	\end{align} 
\end{subequations} 
Utilizing the convolution theory, we can express $G(\kappa)$ by $G(\kappa)=F_1(\kappa)*F_2(\kappa)$, where $F_1(\kappa) = \mathscr{F}[f_1(x)]$ and $F_2(\kappa) = \mathscr{F}[f_2(x)]$.

For the Fourier transform of $f_1(x)$, we have
\begin{equation}
\begin{aligned}
{F}_1(\kappa) &= \frac{1}{\sqrt{2\pi}}\int_{-\infty}^{+\infty}\frac{{-{\rm j} {Z_0} {e^{{\rm j}2\pi \sqrt {{x^2} + {d^2}} /\lambda }}}}{{2\lambda \sqrt {{x^2} + {d^2}} }}e^{{\rm j} \kappa x}{\rm d}x
=A_1+A_2,
\end{aligned}
\end{equation} 
where
$A_1=\frac{1}{\sqrt{2\pi}}\int_{-\infty}^{+\infty}\frac{{-{\rm j} {Z_0} {e^{{\rm j}2\pi \sqrt {{x^2} + {d^2}} /\lambda }}}}{{2\lambda \sqrt {{x^2} + {d^2}} }}{\rm cos}{\kappa x}{\rm d}x$
and
$A_2 = \frac{1}{\sqrt{2\pi}}\int_{-\infty}^{+\infty}\frac{{ {Z_0} {e^{{\rm j}2\pi \sqrt {{x^2} + {d^2}} /\lambda }}}}{{2\lambda \sqrt {{x^2} + {d^2}} }}{\rm sin}{\kappa x}{\rm d}x$.
Since $\frac{{Z_0 {e^{{\rm j}2\pi \sqrt {{x^2} + {d^2}} /\lambda }}}}{{2\lambda \sqrt {{x^2} + {d^2}} }}$ is an even function, it is obvious that $A_2$ equals 0. For $A_1$, we utilize \cite[Eq. (3.876)]{table} to get
\begin{equation}
\begin{aligned}
A_1 = \frac{{-{\rm j} {Z_0} {d^2}}}{{4\pi  \lambda }}\left\{ 
\begin{array} {lcr}
\frac{\pi}{2} {\rm j} {J_0}(dm ) - \frac{\pi}{2} {Y_0}(dm ), & [0 < |\kappa| < \frac{{2\pi }}{\lambda }]\\ 
{K_0}{\rm{(}}dm {\rm{) }}, & [|\kappa| > \frac{{2\pi }}{\lambda }] 
\end{array} 
\right.
\end{aligned}
\label{F1kappa}
\end{equation} 
where $m$ denotes $\sqrt {\left|{{(\frac{{2\pi }}{\lambda })}^2} - {\kappa^2}\right|} $. Then we obtain the closed-form solution of $F_1(\kappa) = A_1$.

For the Fourier transform of $f_2(x)$, we need to derive the Fourier transform of $\frac{1}{{{{({d^2} + {x^2})}^{3/2}}}}$, $\frac{{{x^2}}}{{{{({d^2} + {x^2})}^{3/2}}}}$, $\frac{1}{{{{({d^2} + {x^2})}^2}}}$ and $\frac{{{x^2}}}{{{{({d^2} + {x^2})}^2}}}$.
From \cite[Eq. (3.961)]{table}, we can get
$\mathscr{F}\left[\frac{1}{{{{({d^2} + {x^2})}^{3/2}}}} \right]= \frac{1}{d}\sqrt {\frac{2}{\pi }} \left| \kappa \right|{K_1}(d\left| \kappa \right|)$ and 
$\mathscr{F}\left[\frac{{{x^2}}}{{{{({d^2} + {x^2})}^{3/2}}}}\right]
= \sqrt {\frac{2}{\pi }} {K_0}(d\left| \kappa \right|) - d\sqrt {\frac{2}{\pi }} \left| \kappa \right|{K_1}(d\left| \kappa \right|)$.
For $\frac{1}{{{{({d^2} + {x^2})}^2}}}$, we use residue theorem to get 
\begin{equation}
\begin{aligned}
\mathscr{F}\left[\frac{1}{{{{({d^2} + {x^2})}^2}}}\right] &= 2\pi {\rm j}\,\underset{x = {\rm j}d}{\rm Res}\left[\frac{e^{{\rm j}\kappa x}}{{{{({d^2} + {x^2})}^2}}}\right] \\&= \frac{1}{2d^3}\sqrt {\frac{\pi }{2}} (1 + \left| {d\kappa} \right|){e^{ - \left| {d\kappa} \right|}}.
\end{aligned}
\end{equation}
For $\frac{{{x^2}}}{{{{({d^2} + {x^2})}^2}}}$, we can get
\begin{equation}
\begin{aligned}
\mathscr{F}\left[\frac{{{x^2}}}{{{{({d^2} + {x^2})}^2}}}\right] &  = \mathscr{F}\left[\frac{1}{{{{{d^2} + {x^2}}}}}\right] - {d^2}\mathscr{F}\left[\frac{1}{{{{({d^2} + {x^2})}^2}}}\right]
\\&= \frac{{\sqrt {\frac{\pi }{2}} {e^{ - d\left| \kappa \right|}}}}{d} - \frac{1}{2d}\sqrt {\frac{\pi }{2}} (1 + \left| {d\kappa} \right|){e^{ - \left| {d\kappa} \right|}}
\end{aligned}
\end{equation}

Finally, we can get 
\begin{equation}
\begin{aligned}
F_2(\kappa) &= d{\sqrt {\frac{\pi }{2}} {e^{ - d\left| \kappa \right|}}}+\frac{{\rm j}d\lambda}{2\pi}\sqrt {\frac{2}{\pi }} \left| \kappa \right|{K_1}(d\left| \kappa \right|)\\&~-\frac{{\rm j}\lambda}{\pi}\Big[\sqrt {\frac{2}{\pi }} {K_0}(d\left| \kappa \right|) - d\sqrt {\frac{2}{\pi }} \left| \kappa \right|{K_1}(d\left| \kappa \right|)\Big] \\&~-\left(\frac{\lambda}{2\pi}\right)^2\frac{1}{2d}\sqrt {\frac{\pi }{2}} (1 + \left| {d\kappa} \right|){e^{ - \left| {d\kappa} \right|}}
\\&~+\left(\frac{\lambda}{2\pi}\right)^2\Big[\sqrt {\frac{\pi }{2}}\frac{2{ {e^{ - d\left| \kappa \right|}}}}{d} - \frac{1}{d}\sqrt {\frac{\pi }{2}} (1 + \left| {d\kappa} \right|){e^{ - \left| {d\kappa} \right|}}\Big].
\end{aligned}
\label{F2kappa}
\end{equation}

\ifCLASSOPTIONcaptionsoff
 \newpage
\fi

\footnotesize

\bibliographystyle{IEEEtran}

\bibliography{IEEEabrv,bib}

\end{document}